\documentclass[11pt]{article}
\pdfoutput=1
\usepackage{jheppub}
\usepackage{upgreek}
\usepackage{float, extarrows, tikz-cd}
\usepackage{graphicx}
\usepackage{tabularx,ragged2e}
\usepackage{amssymb}
\usepackage{subcaption}
\usepackage{amsmath,amssymb}
\usepackage{slashed}
\usepackage{caption}
\usepackage{xcolor}
\usepackage{dsfont}
\usepackage{verbatim}
\usepackage{mathtools, xcolor,ytableau, amsfonts,tikz}
\usepackage{graphicx}
\usepackage{physics}
\usepackage{microtype}
\usepackage[nobottomtitles*]{titlesec}
\newcolumntype{C}{>{\Centering\arraybackslash}X}

\numberwithin{equation}{section}

\newcommand{\pd}{\partial}

\newcommand\p{\ensuremath{\partial}}

\def\<{\langle}
\def\>{\rangle}

\author[a,b]{Gabriel Cuomo,}   
\author[a]{Zohar Komargodski}        
              
              \affiliation[a]{Simons Center for Geometry and Physics, SUNY, Stony Brook, NY 11794, USA}      
              \affiliation[b]{C. N. Yang Institute for Theoretical Physics, Stony Brook University, Stony Brook, NY 11794, USA}

\emailAdd{gcuomo@scgp.stonybrook.edu}					 \emailAdd{zkomargodski@scgp.stonybrook.edu}

\begin{document}

\title{Giant Vortices and the Regge Limit}

\abstract{In recent years it has been shown that strongly coupled systems become analytically tractable in the regime of large quantum numbers, such as large spin or large charge. The effective theories that emerge in these two limits are Regge theory and superfluid theory, respectively. Here we make a proposal for a new phase, the ``giant vortex,'' describing an intermediate regime with large spin and charge. The new phase connects superfluid theory with the large-spin expansion.
The giant vortex admits a semi-classical effective theory description with peculiar chiral excitations (moving at the speed of light) and a Fock space of states that is reminiscent of the multi-twist operators in Regge theory, including the leading and daughter Regge trajectories. A similar giant vortex phase appears for Bose-Einstein condensates in a rotating trap, and our results should be applicable in that context as well. 
We show that the transition from the giant vortex to the Regge regime is accompanied by the scaling dimension turning from being larger than to being smaller than the mean field theory value,  i.e. gravity switches from being the weakest force at small AdS distance to being the strongest force at large AdS distance.}

\maketitle

\section{Introduction and Summary}

Conformal Field Theories (CFTs) in $d$ space-time dimensions describe a wide array of physical phenomena, including second order phase transitions in Euclidean $d$ dimensional systems, zero temperature quantum phase transitions in $d-1$ space dimensions,
 and finite temperature phase transitions in quantum systems in $d$ space dimensions. For decades there has been intense work on understanding the spectrum of critical exponents of local operators in CFTs (see~\cite{Poland:2018epd,Chester:2019wfx} for recent reviews). Radial quantization shows that the spectrum of critical exponents is the same as the energy spectrum on $S^{d-1}$. Indeed, choosing the radius of the $S^{d-1}$ to be $R$, we have \begin{equation}\label{radquan}
 E= {\Delta\over R}~,
\end{equation}
 where $E$ is the energy and $\Delta$ is the scaling dimension (critical exponent).
 
The correspondence~\eqref{radquan} allows to think about heavy operators with large $\Delta$, as corresponding to high-energy states. Sometimes, such high energy states admit a useful semi-classical description.
Here we will be interested in $d=3$ theories with a $U(1)$ symmetry with charge $Q$ and we will consider the lowest energy states with some given $Q,J$, where $J$ is the angular momentum.  Taking $Q$ and/or $J$ large we expect that $\Delta$ for the lowest lying operator with the corresponding charge and angular momentum, $\Delta(Q,J)$, would be large as well. Our goal is to understand the behavior of $\Delta(Q,J)$. In terms of the states on $S^2$, we are considering states with large charge and angular momentum, which would be the ground states of the theory in the presence of a chemical potential for the $U(1)$ symmetry and some angular momentum. So the problem we are trying to solve can be equivalently viewed as the search for the (zero temperature) ground state on $S^2$ in the presence of a chemical potential and rotation. The phases of the zero temperature theory in the presence of a chemical potential and rotation could depend on the spatial two-manifold. Here we focus on $S^2$ since this also allows to translate our results to statements about the critical exponents of heavy local operators with spin and charge.
In this paper we focus on the $O(2)$ Wilson-Fisher model in $d=3$, though of course our discussion applies to any other theory with the same semi-classical limits for large $Q,J$. The problem of finding (zero temperature) ground states on $S^2$ in the presence of a chemical potential and rotation is somewhat analogous to the problem of finding the phases of cold atoms in a trap -- therefore, some of our results should apply in that context, as we will explain in more detail below.

Past developments have provided us with quite a lot of information about the case of large $Q$ and relatively small $J$ or large $J$ with relatively small $Q$. Let us briefly summarize those two cases.
For large $Q$ and $J=0$ the ground state on $S^2$ is a conformal superfluid~\cite{Hellerman:2015nra,Monin:2016jmo,Gaume:2020bmp}. One finds\footnote{The superfluid picture is compatible with the bootstrap equations in the macroscopic limit derived in \cite{Jafferis:2017zna}.}
\begin{equation}\label{largeQ}
\Delta=c_1Q^{3/2}+c_2Q^{1/2} - 0.0937+\cdots~,
\end{equation}
with the $\cdots$ standing for corrections which are small for $Q\gg1$ and $c_{1,2}$ are some dimensionless coefficients.\footnote{Their values for the $O(2)$ model are known approximately from numerics (and there exist estimates from the $\varepsilon$ expansion~\cite{Badel:2019oxl} and large $N$~\cite{Alvarez-Gaume:2019biu}), $c_1\simeq 0.337...$ and $c_2=0.266...$~\cite{Banerjee:2017fcx}.} Of course, using the conformal superfluid description, we can compute excitations above the ground state and various other quantities, such as  some of the corrections to the ground state energy in~\eqref{largeQ}.
In the opposite regime, with large $J$ and small or vanishing $Q$ we have the large-spin effective theory~\cite{Alday:2007mf,Komargodski:2012ek,Fitzpatrick:2012yx} where one finds Regge-like behavior with 
\begin{equation}\label{Regge}
\Delta=J+\cdots~,
\end{equation}
with the corrections suppressed for $J\gg 1$. 

Our goal is to chart out the phases of the theory as a function of $J,Q$. This program has been initiated in~\cite{Cuomo:2017vzg}. One finds a rather rich phase diagram. Below we always have $Q\gg1$.
\begin{itemize}
\item $0\leq J\ll Q^{1/2}$: Here one simply adds to the isotropic superfluid a single phonon with spin $J$.\footnote{As we will review, the dispersion relation of phonons is such that a single phonon of angular momentum $J$ is preferable over several phonons each of lower spin,  at least up to $J\sim Q^{1/3}$. In the domain $Q^{1/3}\lesssim J\ll Q^{1/2}$ whether one phonon is preferable over several is dependent on small corrections to the dispersion relation of phonons.  The energy of a multiphonon state differs from that of a single phonon with the same angular momentum only in the $O(1)$ term, thus eq. \eqref{eq_sflu} remains true to leading order in $J$.
\label{footnote_dis}} The effective field theory of the superfluid has cutoff at distance scales $Q^{-1/2}R$, which implies that we can trust phonons with $J\ll Q^{1/2}$. 
\begin{equation}\label{eq_sflu}
\Delta = c_1Q^{3/2}+c_2Q^{1/2}+{1\over \sqrt 2}\sqrt{J(J+1)} - 0.09372+\cdots~.
\end{equation}
\item $Q^{1/2} \ll J\ll Q$: As is familiar from rotating superfluids, at some point it becomes favorable to form vortices. This regime is described by a single vortex anti-vortex pair (rotating around the sphere) and as the spin is increased, the vortices move farther apart until one is at the north pole and the other is in the south pole. The spin cannot be increased further with just a vortex anti-vortex pair.
\begin{equation}\label{eq_2vortices}
\Delta = c_1Q^{3/2}+{\sqrt Q\over 6c_1}\log{J(J+1)\over Q}+\cdots~.
\end{equation}
Note the logarithmic enhancement in this regime compared to the $c_2$ term in~\eqref{largeQ}. The vortex masses introduce additional $O(Q^{1/2})$ corrections to eq.  \eqref{eq_2vortices} which depend upon a Wilson coefficient independent from $c_2$ introduced in eq. \eqref{eq_sflu}.
\item $Q \ll J\ll Q^{3/2}$: Here many vortices (and anti-vortices) appear -- in fact their density on the $S^2$ is proportional to $\cos(\theta)$ and they rotate around the $S^2$ as if they form a rigid body -- the angular velocity of the vortices is constant (in the non-relativistic approximation). The equation of state in this regime is  
\begin{equation}\label{eq_many_vortices}
\Delta = c_1Q^{3/2}+{1\over 2c_1} {J^2\over Q^{3/2}}+\cdots ~.
\end{equation}
For $J\sim Q^{3/2}$ the rotation speed becomes relativistic,  
the density of vortices grows and when the rotation speed exceeds the speed of sound presumably a superradiant instability develops \cite{Bekenstein:1998nt}, leading to a new phase at $J\sim Q^{3/2}$. 

\item $Q^{3/2} \ll J\ll Q^{2}$: We refer to the effective theory in this regime as the ``giant vortex.''  The superfluid now does not extend throughout the $S^2$ but, instead, is concentrated in some domain $[\pi/2-\delta,\pi/2+\delta]$ around the equator. The superfluid in that domain is spinning relativistically fast. 
The equation of state is
\begin{equation}\label{eq_giant_vortex}
\Delta = J +{9c_1^2\over 4\pi} {Q^3\over J}+\cdots~.
\end{equation}
When $J\sim Q^2$ the domain becomes very narrow and the effective field theory breaks down. 
We see that, remarkably, the energy becomes linear in $J$ in this giant vortex regime (with coefficient exactly 1), as in the Regge regime, which follows next. Additionally, the second term becomes of order $Q$ near the boundary of the regime of validity of the giant vortex, i.e. $Q^3/J\sim Q$ for $J\sim Q^2$. This is again, exactly as in the Regge regime. 
\item $Q^{2} \ll J$: Here we reach the Regge regime, where there are $Q$ ``partons'' spinning around the equator. This description of the partons breaks down for $J\sim Q^2$ since, as we will see, they begin to strongly interact and indeed, presumably, as $J$ is lowered further, they collapse into the giant vortex state.  The scaling dimension obeys
\begin{equation}\label{Reggeint}
\Delta=J+\Delta_\Phi Q+\cdots~,
\end{equation}
where $\Delta_\Phi$ is the dimension of the $Q=1$, spinless parton.  To leading order $\Delta$ is insensitive to how the $J$ derivatives are distributed among the $Q$ partons, leading to a large approximate degeneracy in the large spin expansion.
\end{itemize} 

\begin{figure}[t]
\begin{center}
\includegraphics[scale=0.35,trim= 0 0 2em 5em]{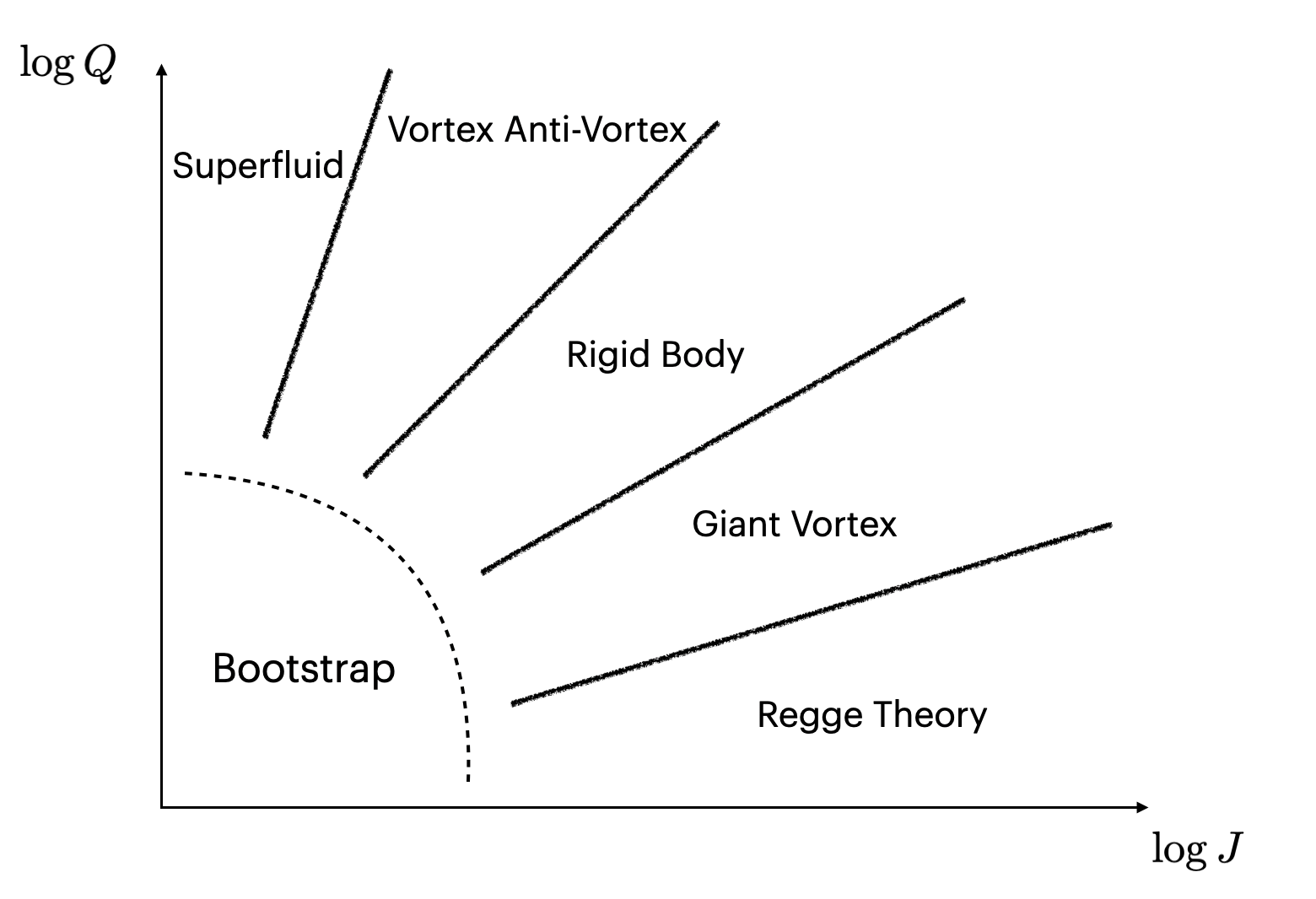}\caption{Phase-diagram for the ground-state on $\mathbb{R}\times S^2$; the small $Q$ and $J$ regime is accessible via the numerical bootstrap.}\label{phases}
\end{center}
\end{figure}

The phases with $J\ll Q^{3/2}$ were already described in~\cite{Cuomo:2017vzg}. One main new result is the giant vortex phase, which we conjecture to be the ground state after for $Q^2\gg J\gg Q^{3/2}$.  The giant vortex interpolates between the superfluid description and the Regge limit.
Fig.~\ref{phases} summarizes the semi-classical phases we discuss at large $Q,J$.\footnote{ The phases outlined above are truly the ground states at fixed $Q,J$. It is not possible to get lower-energy states with these quantum numbers from descendants of states with smaller $J$. For instance, descendants of the homogeneous superfluid are more energetic than the vortex anti-vortex pair, the rigid body, the giant vortex, and the Regge partons.}

We encounter surprising aspects of the effective theory of the giant vortex phase. The fluctuations around the giant vortex are in a sense {\it chiral}; the fluctuations form a Fock space which is very similar to the Fock space of multi-twist operators in the large-spin effective theory! The Fock space of the giant vortex includes both the leading twist Regge trajectory as well as the daughter trajectories, labeled by an integer.
The giant vortex phase interpolates between the vortex rigid body and the Regge limit in a very compelling fashion and has qualitative (and some quantitative) similarities with the large spin limit.  For instance, both in the large spin limit and in the giant vortex, there is an approximate infinite degeneracy of states. In the large spin limit, as we remarked above, this is due to the insensitivity of the scaling dimension to the distribution of the $J$ derivatives, while in the giant vortex phase this is due to the peculiar chiral excitations which can either raise or lower the energy.  In addition, the giant vortex phase is also motivated by the study of rotating traps, see e.g.~\cite{kasamatsu2002giant,Fischer:2003zz,fetter2005rapid,fu2006transition, Fetter:2009zz,guo2020supersonic}, and it was also discussed in the context of holography~\cite{Su:2022ysc}. Our results about the peculiar Fock space of fluctuations around the giant vortex should be applicable in this wider context.\footnote{Technically similar solitons were also recently analyzed in \cite{Penin:2020cxj,Penin:2021xgr}.}

From the point of view of the large-spin effective theory, the $Q$ partons experience strong interaction at $J\sim Q^2$ while they are weakly coupled at  $J\gg Q^2$. 
Therefore, the giant-vortex may be viewed as a proposal for the new ground state, replacing the $Q$-parton state for $J\sim Q^2$. The strong interaction is due to the energy momentum and current exchange, and hence, it might be  possible to re-sum those exchanges and confirm that the new ground state is the giant vortex.  In the AdS$_4$ analog of the CFT, a multitrace large spin operator with $J\gg Q^2$ consists of $Q$ \emph{blobs} orbiting at superhorizon distance $d\sim R_{AdS}\log(J/Q)\gg R_{AdS}$ from one another \cite{Komargodski:2012ek,Fitzpatrick:2012yx,Fitzpatrick:2014vua}. The leading interaction arises from the gravitational and the electric forces between the blobs, corresponding, respectively, to energy momentum and current exchanges in the CFT. Consistency with our picture requires the gravitational attraction to dominate over the electromagnetic repulsion, so that this configuration may plausibly collapse to a coherent superfluid state for lower $J$. This is indeed what we find, as we further discuss below.

\paragraph{Connection to the Weak Gravity Conjecture}
We can think of $\Delta(Q,J)$ as the energy of a composite of $Q$ particles in a state with angular momentum $J$. 
In an AdS language the gravitational force is always attractive while the electric forces are repulsive since like charge repel. Therefore, if gravity is the weakest force \cite{Arkani-Hamed:2006emk} (see~\cite{Palti:2019pca} for a general review)  one should expect $\Delta(Q,J)$ to be above the mean field theory value $\Delta_{{\rm free}}(Q,J)=J+\Delta_\Phi Q$.  This is manifestly true for superfluid and vortices, since $\Delta(Q,J)\sim Q^{3/2}$ and $J\ll Q^{3/2}$. In the giant vortex phase this is still true, but the inequality relies on the sign of the subleading term in eq. \eqref{eq_giant_vortex}; this means that the repulsion becomes parametrically smaller for $J\gg Q^{3/2}$. In particular, the subleading term in eq. \eqref{eq_giant_vortex} dangerously becomes $O(Q)$ as the angular momentum approaches the boundary of the window of validity, $J\rightarrow Q^2$, from below.  
Eventually, in the Regge regime, in the approximation~\eqref{Reggeint} we have vanishing binding energy since to leading order $\Delta(Q,J)=\Delta_{\rm{free}}(Q,J)$. In the main text we compute the leading correction to~\eqref{Reggeint} in the $O(2)$ model and find that in the Regge regime $\Delta(Q,J)<\Delta_{\rm {free}}(Q,J)$.
In an AdS language, one could say that at large separation in AdS the gravitational force among $Q$ charged particles is stronger than the electric repulsion between them, while for small separation the opposite is true.
As $J$ is increased, $\Delta(Q,J)\to \Delta_{\rm {free}}(Q,J)$ from below.\footnote{In fact, our computation in the main text shows that $\Delta(Q,J)<\Delta_{\rm {free}}(Q,J)$ for large enough $J$ also for $Q\sim1$, not just for large $Q$; this is in agreement with the analysis of  \cite{Li:2015rfa,Liu:2020tpf} for $Q=2$.} In the special case $J=0$, $\Delta(Q,J)>\Delta_{\rm {free}}(Q,J)$ translates to $\Delta(Q,0)>Q\Delta_\Phi$ which follows from the conjectured convexity~\cite{Aharony:2021mpc,Antipin:2021rsh,Moser:2021bes,Palti:2022unw}. Here we see that for larger separation in AdS, the inequality is reversed. It would be interesting to understand how general this sign reversal is.

\paragraph{Outline}
The outline of the paper is as follows. In section 2 we review the large $Q$ limit with small $J\ll Q^{1/2}$. In section 3 we review the vortex anti-vortex pair and the rigid body phase. In section 4 we introduce the giant vortex phase, analyze and quantize the fluctuations and compare with the Regge limit (which we also briefly review in this section). Finally, in section 5 we discuss several open questions. Appendices cover some technical details.

\section{The Large Charge Limit}\label{sec_largeQ}

It is natural to assume that the lightest scalar operator with $U(1)$ charge $Q$ corresponds to a state with homogeneous charge density on $S^2\times \mathbb{R}$.  For $Q\gg 1$ the charge density sets a dimensionful scale $\mu\sim \sqrt{Q}/R$ much larger than the compactification scale $1/R$, where $R$ is the sphere radius. The CFT is thus  in a ``condensed matter" phase. The simplest possibility is that this phase is a superfluid one \cite{Nicolis:2015sra,Hellerman:2015nra}.\footnote{Another consistent possible phase is a Fermi surface~\cite{Alberte:2020eil,Komargodski:2021zzy}.} This in turn admits a simple effective field theory (EFT) description in terms of a $U(1)$ Goldstone field, with all other ``radial" degrees of freedom separated by a large gap $\sim\mu$.

We therefore conclude that the limit of large $Q$ and small or vanishing $J$ is described by a Weyl invariant action for a $2\pi$ periodic scalar field~\cite{Hellerman:2015nra}
\begin{equation}\label{Hthe}
S=\alpha\int d^3x\sqrt g \left[ |\partial\chi|^{3}+\ldots\right]
~\,,
\end{equation}
where $|\pd\chi|=|\pd_\mu\chi\pd^\mu\chi|^{1/2}$ and the $\ldots$ stand for higher derivative terms.
Of course we are ultimately interested in this theory on $S^2\times \mathbb{R}$ since then the spectrum of energies is the same as the spectrum of local operators~\eqref{radquan}. 
The metric is $ds^2=dt^2-R^2(d\theta^2+\sin^2\theta d\phi^2)$. The theory~\eqref{Hthe} is singular if expanded around the trivial configuration of $\chi$, instead, we expand around $\chi=\mu t $ and denote the fluctuations $\varphi$. Choosing $\mu$ appropriately, we can therefore think about the theory~\eqref{Hthe} as the description of the small fluctuations around the minimal operator of charge $Q$. Alternatively, it can be viewed as describing the ground state of  the CFT on $S^2\times \mathbb{R}$ in the presence of a chemical potential for the $U(1)$ symmetry. Finally, since we are expanding around $\chi=\mu t$ we must contemplate various higher derivative terms suppressed by the scale $\mu$. Such terms must be Weyl invariant and they were discussed in~\cite{Hellerman:2015nra,Monin:2016jmo}.  For future reference we report the two terms which arise at order $O(\pd^2/\mu^2)$:\footnote{We chose the operators multiplied by $\alpha_2$ and $\alpha_3$ to be exactly Weyl invariant, and not only up to a boundary term; see~\cite{Creminelli:2022onn} for a discussion of the constraints from unitarity on the values of $\alpha_2$ and $\alpha_3$.}
\begin{equation}\label{eq_termsNLO}
\begin{split}
S&\supset \alpha_2\int d^dx\sqrt{g} |\p\chi|^3\left[ \frac{\mathcal{R}}{|\p\chi|^2}
-8\frac{\nabla^2\left(|\p\chi|^{1/2}\right)}{|\p\chi|^{5/2}}\right]\\
&+\alpha_3\int d^dx\sqrt{g} |\p\chi|^3\left[\mathcal{R}_{\mu\nu}\frac{\p^\mu\chi\p^\nu\chi}{|\p\chi|^4}
-\frac{\nabla^2|\p\chi|}{|\p\chi|^{3}}
+\frac{\p^\mu\chi\p^\nu\chi\nabla_\mu\nabla_\nu\left(|\p\chi|^{-1}\right)}{|\p\chi|^{3}}
\right]\,,
\end{split}
\end{equation}
where $\mathcal{R}^\mu_{\;\nu\rho\sigma}$ is the Riemann tensor.
The scale $\mu$ is therefore the natural cutoff for the EFT.\footnote{While in strongly coupled models the Wilson coefficients $\alpha_i$ are $O(1)$ (or more precisely scale with inverse powers of $(4\pi)$ according to generalized dimensional analysis \cite{Georgi:1992dw}) and $\mu\sim\sqrt{Q}/R$, in weakly coupled models $\mu$ may be parametrically smaller than $\sqrt{Q}/R$ and Wilson coefficients may have nongeneric scaling. We refer to \cite{Cuomo:2017vzg} for a general discussion of the derivative expansion in those cases and to \cite{delaFuente:2018qwv,Alvarez-Gaume:2019biu,Badel:2019oxl} for some examples.}

To find the relationship between $Q$ and $\mu$ and to find the energy of the configuration $\chi=\mu t$ (as well as for future reference) we write the current and energy-momentum tensor which follow from eq. \eqref{Hthe}:
\begin{equation}\label{current} 
j_\mu= 3\alpha |\partial \chi | \partial_\mu\chi~,
\end{equation} 
\begin{equation}\label{emtensor} 
T_{\mu\nu}=\alpha \left( 3|\partial \chi |\partial_\mu\chi\partial_\nu\chi-g_{\mu\nu}|\partial \chi|^3 \right)~.
\end{equation} 
Note that the energy-momentum tensor is  traceless. Remembering that the volume of the two-sphere is $4\pi R^2$ we see that the total charge is $Q=12\pi R^2\alpha\mu^2$ and the total energy is 
$E=8\pi R^2 \alpha \mu^3 $ and hence we find that 
\begin{equation}\label{lcharge}
\Delta = RE= {1\over 3^{3/2}\sqrt{\pi \alpha}}Q^{3/2}+\ldots
\end{equation}
Therefore the coefficient $c_1$ that appeared, for instance, in~\eqref{largeQ}, satisfies $c_1={1\over 3^{3/2}\sqrt{\pi \alpha}}$ in terms of the coefficient $\alpha$ that appeared in the action~\eqref{Hthe}. The coefficient $c_2$ arises from terms \eqref{eq_termsNLO} which are suppressed by the cutoff $\mu$.

The fluctuations $\chi=\mu t + \varphi$ to second order have an action\footnote{Indices $i,j,\ldots$ are contracted, raised and lowered with the Euclidean metric $g_{ij}$.}
\begin{equation}\label{fluc}
S_{\rm fluctuations}= 
3\mu^2\alpha\int d^3x\sqrt{g}\, \partial_0\varphi+
3\mu \alpha \int d^3x\sqrt g  \left[ (\partial_0\varphi)^2 -\frac 12 (\partial_i\varphi)^2 \right]~.
 \end{equation}
The first term is a total derivative -- we keep it for a reason that will become clear in the next section. Expanding the fluctuations in spherical harmonic modes, we see that the total derivative term vanishes for all but the constant mode. So let us first quantize the non-constant modes. They obey the dispersion relation ($J\neq 0$) 
\begin{equation}\label{disrel_phonon}
\omega^2=\frac {1}{2 R^2} J(J+1)~.
\end{equation} 
The corresponding modes are phonons of the superfluid. Their speed of propagation is $1/\sqrt 2$ of the speed of light, which is not surprising given that this is the speed of sound in an isotropic conformally invariant medium in 2+1 dimensions. In particular, the descendants correspond to states involving a number $n > 0$ of spin one quanta, each increasing the energy by $\omega=1/R$. The zero modes of $\varphi$ instead allow interpolating between states with different charge .

Since the cutoff of our theory is $\mu$, we cannot excite too many phonons -- we see from the dispersion relation that we should only trust phonons below $J \sim \mu R$ which is why this description is valid up until $0<J\ll\sqrt Q$. Note that $\omega(J)$ is a sub-additive function, i.e. $\omega(J_1+J_2)\leq \omega(J_1)+\omega(J_2)$. This means that to minimize the energy for a given $Q$ and $J$ it is always beneficial to excite exactly one phonon of spin $J$.\footnote{In general there is a contribution $\sim J^3/(R^2\mu^2)$ to the dispersion relation \eqref{disrel_phonon} which arises from the higher derivative operators \eqref{eq_termsNLO}.  For $J\gtrsim(R\mu)^{2/3}\sim Q^{1/3}$ the subadditivity of $\omega(J)$ depends on the sign of this contribution and multi-phonon states might be preferred over single-particle ones \cite{Firat}.
}
The detailed map between the Fock space of phonons and the primary operators of charge $Q$ (with various derivatives inserted) is discussed in~\cite{Badel:2022fya} for some weakly coupled theories.

\section{Vortices}\label{sec_vortices}

\subsection{Vortex Dynamics}

Since the phonons cannot describe states with $J$ exceeding $\sqrt Q$, we must now switch to a description that also contains vortices.\footnote{The important role of vortices in superfluids was recognized long ago~\cite{onsager1949statistical,feynman1955chapter}. For the study of vortices in the wave function of particles in a rotating trap see e.g. also~\cite{castin1999bose,aftalion2001vortices,aftalion2007vortices} in addition to the references already mentioned in the introduction.} Below we review some aspects of the dynamics of vortices.\footnote{The following analysis provides an alternative (equivalent) treatment to the one in \cite{Horn:2015zna,Cuomo:2017vzg}, bypassing the explicit dualization of the action \eqref{Hthe}.}

In the previous section that $\chi$ is compact was only important for the quantization of $Q$ itself.
However, our spatial configurations of $\chi$ were single valued, regular functions on $S^2$ (such configurations are called irrotational). In this section we explore the configurations of $\chi$ for which the spatial configuration is not single valued. A vortex with core $x_0(t)$ is some configuration $\varphi(t, x)$ that it is not single valued when $x$ is rotated around the core $x_0(t)$. We say that there are $q$ units of vorticity if $\varphi(t, x)$ shifts by $2\pi q$ as we go around the core (of course $q\in\mathbb{Z}$). 

Vortices are not finite energy configurations in~\eqref{Hthe}. Indeed, from the quadratic term $\sim \int dt d^2x \sqrt g (\partial_i\varphi)^2$ in~\eqref{fluc}, we see that near the core of the vortex, in polar coordinates around the core, the energy of a vortex behaves like $\int {dr\over r} \sim \log\left({L \over a}\right)$ where $a$ is a short distance cutoff and $L$ is a long distance scale (which will be soon identified with the location of an anti-vortex).

The energy of any configuration containing vortices would therefore always be infinite in the limit $a\to 0$, but if one assumes that in the full theory vortices do have a finite energy (as would be the case in the $O(2)$ model, where at the core of the vortex the $O(2)$ order parameter vanishes, and by dimensional analysis the mass of a vortex must be proportional to $\mu$), these are the relevant configurations to consider for minimizing the energy as a function of $Q,J$ with $J\gg Q^{1/2}$. The calculable effects below depend only logarithmically on the cutoff $a$ and hence it would not be an error within our approximation to identify the cutoff $a$ with $\mu^{-1}$. We therefore  imagine for simplicity that the mass and inverse radius of vortices are both approximately $\mu$, since there is no other scale.

The action for a vortex moving along a prescribed trajectory  receives contributions from the quadratic terms in the fluctuations but also from the linear term in~\eqref{fluc}. To analyze the contribution of the linear term $3\alpha \mu^2\int d^3x \sqrt g \partial_0 \varphi$ to the action of a vortex at $x_0(t)$,  we observe that this term fails to be a total derivative only around the location of the vortex. Therefore, we pick normal coordinates $y^i$ in the vicinity of the vortex, in terms of which $\varphi=\varphi(y^i-y_0^i(t))$ very close to the vortex core. The linear term in the action thus becomes 
\begin{equation}
3\alpha \mu^2\int d^3x \sqrt g \partial_0 \varphi=3\alpha \mu^2\int dt d^2y{\partial\varphi\over\partial y^i}  \dot{y}_0^i(t)~.
\end{equation}
We only need to evaluate the integral $I_i=\int d^2y{\partial\varphi\over\partial y^i}$. 
To evaluate this integral we use the local definition of $q$ units of vorticity
\begin{equation}\label{eq_V_ij}
{\mathcal V}_{ij}=\partial_i{\partial\varphi\over\partial y^j}-\partial_j{\partial\varphi\over\partial y^i}= 2\pi q\epsilon_{ij}\delta^{2}(y-y_0(t))\,,
\end{equation}
with $\epsilon_{ij}$ being the usual $\pm 1$ symbol in any coordinate system.
With simple manipulations we then find\footnote{
To obtain eq. \eqref{eq_I_i} we take another derivative with respect to $y_0$ and contract with the epsilon tensor: $\epsilon^{ki}\partial_{y_0^k}I_i=-\epsilon^{ki}\int d^2y\partial_{k}{\partial\varphi\over\partial y^i}$. Now we use the property~\eqref{eq_V_ij}. Thus we find $\epsilon^{ki}\partial_{y_0^k}I_i=-2\pi q$ from which we infer that 
$I_i=\pi q \epsilon_{ij} y_0^j$.}
\begin{equation}\label{eq_I_i}
I_i=\int d^2y{\partial\varphi\over\partial y^i}=\pi q \epsilon_{ik} y_0^k\,.
\end{equation}
Plugging this result into the action of the vortex we therefore find the following term in the action 
\begin{equation}
S\supset  3\alpha \mu^2\int dt d^2y{\partial\varphi\over\partial y^i}  \dot{y}_0^i(t)=3\alpha \mu^2 \pi q \int dt \epsilon_{ik} y_0^k \dot{y}_0^i(t)\,.
\end{equation}
This action for the vortex core is very familiar from the motion of a classical particle in a transverse constant magnetic field (leading to the Lorentz force). To make the analogy explicit, we introduce the effective magnetic field $\partial_{  i} A_{j}-\partial_j A_i=-6\alpha\mu^2\pi \epsilon_{ij}\sqrt{g}$. We therefore see that each vortex has the following term in its worldline action 
\begin{equation}\label{lorentz} 
S\supset q\int dt  A_i(y(t)) \dot{y}^i~,\qquad \partial_{  i} A_{j}-\pd_j A_i=-6\alpha\mu^2\pi \epsilon_{ij}\sqrt g =-\frac{Q}{2R^2}\epsilon_{ij}\sqrt{g}~.
\end{equation} 
Eq. \eqref{lorentz} allows to lift the worldline action to one in general coordinates. Note that this is a properly quantized magnetic field on $S^2$, with exactly $Q$ units of the minimal magnetic field.
The fact that this effective magnetic field is properly quantized guarantees that no Dirac string is needed in the vector potential $A_i(y(t))$ and hence the vortices move on the two-sphere in an $SO(3)$ invariant fashion.

The other contribution to the action of the vortices comes from the quadratic terms in the fluctuations. We will assume that the vortices move at velocity much smaller than the speed of light so the quadratic time derivative terms do not matter. From the quadratic space derivative terms, we have already argued that there is a logarithmic interaction -- let us compute it in detail now, starting from the action $-{3\mu\alpha\over 2} \int dt d^2x \sqrt g (\partial_i\varphi)^2$. For vortices with locations $x_\alpha(t)$ where the index $\alpha$ labels the vortices we have the equation of motion $\Delta_{S^2} \varphi=0$ away from the vortices (we assume the vortices are slowly moving so we only solve the space-like Poisson equation) which is solved by
\begin{equation}\label{vopro}
 \partial_i  \varphi =2\pi \sqrt g \epsilon_{ij} \sum_\alpha q_\alpha \partial^j G(x,x_\alpha) ~, 
 \end{equation}
with $q_\alpha$ the vorticity and $G$ the standard Green's function on the sphere:
\begin{equation}
G(x,y)=-\frac{1}{4\pi}\log\left(\hat{n}_x-\hat{n}_y\right)^2\,.
\end{equation}
Here $\hat{n}$ is unit normalized three-vector describing the embedding of the $2$-sphere in $\mathbb{R}^3$; $R^2(\hat{n}_x-\hat{n}_y)^2\equiv L_{xy}^2$ is the \emph{chordal} distance between the points $x$ and $y$ on $S^2$.

Using eq.~\eqref{vopro} in the action and after a little bit of algebra we find the static component of the logarithmic interaction between the vortices \begin{equation}\label{eq_voraction_pre}
S\supset 3\pi \mu\alpha\int dt\left[\sum_{\alpha\neq \beta} q_\alpha q_\beta \log(\mu L_{\alpha\beta})+\text{const.}\right]\,.
\end{equation}
The cutoff $\mu$ appears explicitly in the static potential between vortices, which accounts for the self energy as well, due to the constraint that the total number of vortices vanishes. The constant term in eq. \eqref{eq_voraction_pre} is interpreted as the contribution of the vortex masses and it is an independent Wilson coefficient within EFT. (More precisely, this independent Wilson coefficient is due to the ratio of the vortex mass and the scale $\mu$, which is a model dependent coefficient, expected to be $\sim 1$ in the $O(2)$ model.)

We are now ready to summarize the action of the vortices: 
\begin{equation}\label{voraction}
S_{vortices}=\int dt  \left[\sum_\alpha q_\alpha A_{i}(x^i_\alpha) \dot x^i_\alpha(t)+
3\pi\mu\alpha \sum_{\alpha\neq \beta} q_\alpha q_\beta \log\left(\mu L_{\alpha\beta}\right)\right]~,
\end{equation}
where the gauge field satisfies $\partial_{ i} A_{ j}-\pd_j A_i=-6\alpha\mu^2\pi  \epsilon_{ij}\sqrt g=Q/(2R^2)\epsilon_{ij}\sqrt{g}$ and we neglected the masses of the vortices. The kinetic term $\sim\mu \int dt\sum_\alpha\dot x^2_\alpha$ is negligible for slowly moving vortices $\dot x\ll 1$; quantum-mechanically, neglecting the kinetic term is equivalent to restricting the dynamics to the lowest Landau level \cite{integrateLandau1,integrateLandau2,integrateLandau3,jackiw1,jackiw2}.

The energy stored by the vortices, on top of the ground state energy $8\pi R^2\alpha\mu^3$,
is independent of the magnetic field and is just given by the electrostatic potential:
\begin{equation}\label{vor_energy}
 E_{vortices}=-3\pi\mu\alpha \sum_{\alpha\neq \beta} q_\alpha q_\beta \log\left(\mu L_{\alpha\beta}\right)~.
\end{equation}
The charge is not sensitive to the vortices and it is given as before by $Q=12\pi R^2\alpha\mu^2$. 

The angular momentum is sensitive to the vortices. Let us compute the angular momentum of a static $q$ vortex at the north pole. From~\eqref{emtensor} we find $T_{0\phi}= 3\alpha \mu^2 q$, therefore a single vortex contributes to $J_z$ by $\frac12\int d^2x T_{0\phi} =6\pi R^2\alpha \mu^2q$. We divided by 2 since the configuration $\varphi=q\phi$ has a $q$ vortex in the north pole and a $q$ anti-vortex in the south pole, each of which is contributing a half to the total angular momentum. More generally, the angular momentum is given by 
\begin{equation}\label{angmo}
\vec J = 6\pi R^2\mu^2 \alpha \sum_\alpha q_\alpha\hat n_\alpha~,\end{equation} 
where $\hat{n}_\alpha$ is the unit vector pointing from the center of the sphere to the location of the vortex. Of course, the angular momentum also receives contributions from the motion of vortices but those are again negligible for slowly moving vortices.\footnote{The fact that a single vortex has angular momentum is of course analogous to the angular momentum in the monopole-charge system~\cite{Shnir:2005vvi}. In the presence of many vortices, the angular momentum is additive~\eqref{angmo} since the equation of motion for the fluctuations around $\chi=\mu t$ is linear to leading order and hence the profile of $\varphi$ is a superposition over all the vortices~\eqref{vopro}. }

\subsection{Results}

Consider a vortex anti-vortex pair, each with unit vorticity.  When the vortices are antipodally placed, as in~\eqref{antipodal}, they are static. More generally, we have to balance the Lorentz force, which is the first term in~\eqref{voraction} with the electric potential, which is the second term in~\eqref{voraction}.  Vortices therefore move via drift motion with velocity $v\sim1/(\mu L)$ around the sphere, where $L$ is the relative distance.  The energy and angular momentum of the configuration are
\begin{equation}\label{antipodal}
\Delta/R=8\pi R^2\alpha\mu^3+6\pi \mu \alpha \log(\mu L)+\ldots\,,\qquad
J_z=6\pi R^2\mu^2\alpha L=\frac{Q}{2}\frac{L}{R}\,.
\end{equation}
The result \eqref{eq_2vortices} follows upon rewriting \eqref{antipodal} in terms of $Q$ (the $+1$ in $J(J+1)$ follows from quantization \cite{Cuomo:2017vzg}).
Eq. \eqref{antipodal} receives $O(\mu)$ corrections both from the higher derivative terms \eqref{eq_termsNLO} and from the vortex masses. By moving the vortices farther apart we can increase the angular momentum while keeping the charge fixed. In this way we can increase the angular momentum until we reach $J_z=Q$, when the vortices are antipodal. On the lower end of the regime of validity of the vortex anti-vortex system, for $J_z\lesssim \sqrt{Q}$, the vortices are nearby and they become relativistic so our EFT breaks down. Indeed, there are no vortices for $J_z\lesssim \sqrt{Q}$, as we have seen, phonons are sufficient in this regime. In summary, the vortex anti-vortex configuration is well suited to be the ground state of the system for $\sqrt Q\ll J\leq Q$.

The angular momentum can be increased seemingly as much as we like if we just keep increasing  the vorticity $q$. But this is not the ground state since, for instance, a 2-vortex is generally going to be unstable towards breaking up to two single vortices.\footnote{This is because the self-energy term scales quadratically with the vorticity. We can therefore clearly decrease the energy by breaking up a $2$-vortex and separating the 1-vortices and 1-anti-vortices. 
This can be done without violating angular momentum conservation. Neglecting the kinetic term, angular momentum~\eqref{angmo}  can be conserved by simply moving the center of mass of the vortex binary slightly farther from the center of mass of the anti-vortex binary. The splitting of the 2 vortex into a binary of 1-vortices slightly decreases the angular momentum and this can be compensated for by moving the binaries a little farther apart, which is possible as long as the original 2-vortex 2-anti-vortex pair is not antipodal. In the latter case we conjecture that it is still possible for the 2 vortex and 2 anti-vortex apart to break up into relativistic binaries (see also the discussion in \cite{Garaud:2020qeq}). More broadly, in 2+1 dimensional theories with $U(1)$ symmetry and one-form symmetry, there could be additional selection rules for which vortices exist and which do not and the conclusions could thus change.}
Given these arguments we now look for configurations that can take us beyond $Q\sim J$. We first consider configurations that are made out of 1-vortices and 1-anti-vortices only.

For sufficiently large $J$ we may approximate the vortex distribution with a continuous function $\rho(x)$.
The vortex contribution to the energy \eqref{vor_energy} thus reads:
\begin{equation}\label{rigidboE} 
E_{vortices}=
6\pi^2\mu\alpha\int d^2x\sqrt g d^2x'
\sqrt {g'} \rho(x)\rho(x')G\left(x,x'\right)\,,
\end{equation} 
while the angular momentum is 
\begin{equation}\label{rigidboAn}
\vec J=6\pi\alpha\mu^2R^2\int d^2x\sqrt g \rho(x)\hat{ n}(x)~.
\end{equation}

To minimize the energy at fixed angular momentum we assume $J_x=J_y=0$ and consider the functional $E_{vortices}+\lambda J_z$, from which we obtain the following minimum condition
 \begin{equation}\label{eomGMi}
\int d^2x'\sqrt {g'}\rho(x')G\left(x',x\right)= \frac{\lambda}{2}\mu R^2\cos(\theta)\,,
\end{equation}
where $\lambda$ is a Lagrange multiplier.
The first term on the left hand side of~\eqref{eomGMi} is the electric potential due to charges with density $\rho$ and the second term on the right hand side states that this electric potential is proportional to $\cos\theta$. The equation~\eqref{eomGMi} is of course solved by acting on both sides with $\Delta_{S^2}$ in the $x$ coordinates.  Expressing $\lambda$ in terms of $J_z$, we find the following result for the vortex density
\begin{equation}\label{vortden}
\rho={J_z\over 8\pi^2 \alpha \mu^2R^4} \cos(\theta) =\frac{3 J_z}{2\pi Q R^2}\cos\theta\,.
\end{equation}
Eq. \eqref{vortden} corresponds to a velocity profile of a rigid body with a spherical shape,  explicitly $v_\phi=j_\phi/j_0 \simeq \sin^2\theta J/(cQ^{3/2})$ (therefore $v^\phi$ is constant and hence the angular velocity is $\theta$ independent, as for a rigid body). The energy of the rigid body as a function of $Q,J$ is inferred from~\eqref{rigidboE}:
\begin{equation}\label{finalErigid}
E={1\over 3^{3/2}R\sqrt{\pi \alpha}}Q^{3/2}+{3^{3/2} \sqrt{\pi\alpha} \over 2R} {J^2\over Q^{3/2}}~.
\end{equation}
We see that for $J\gg Q$ the second term in~\eqref{finalErigid} parametrically dominates over the term depending on the Wilson coefficient $c_2$ in~\eqref{largeQ}.  Additionally the vortex masses contribute to order $\sim\rho\times \mu\sim J/\sqrt{Q}$, which is subleading for $J\gg Q$.\footnote{ The vortex mass contribution can be read from~\eqref{eq_voraction_pre}. While the constant piece is model dependent, the logarithmic self energy piece $-3\pi \mu\alpha\sum_{\alpha\neq\beta}q_\alpha q_\beta\log(\mu)=3\pi \mu\alpha\sum_{\alpha}q^2_\alpha \log(\mu)$ is calculable. The microscopic realization of the rotating rigid body only has 1-vortices and 1-anti vortices and therefore, if the total number of vortices and anti-vortices is denoted by $\mathcal N$, then we find the total self energy $3\pi \mu\alpha{\mathcal N} \log(\mu)$. For the rigid body configuration ${\mathcal N}={3 J_z\over Q}$ and therefore the self energy is 
\begin{equation}
\frac{3\sqrt{3\pi \alpha}}{ 4R} \frac{J_z}{\sqrt{Q}} \log Q=\frac{J_z}{4c_1\sqrt{Q}}\log Q~,
\end{equation}
where we kept the large logarithm which is unaffected by the unknown Wilson coefficients. This is parametrically smaller than~\eqref{finalErigid} since $J\gg Q$ in the rigid body phase. \label{footnote_mass}
} We notice that, since the number of vortices is quantized, the derivation leading to eq. \eqref{finalErigid} only holds for $3J/Q\in\mathds{Z}$ (see footnote~\ref{footnote_mass}). It would be interesting to generalize our treatment to more general values of $J$.

To remain within the non-relativistic regime we must require that $J\ll Q^{3/2}$, since as $J$ comes close to $Q^{3/2}$ the rotation velocity becomes relativistic and our treatment of the problem needs to be revisited.
The more general case, including the first nontrivial relativistic correction, is worked out in appendix~\ref{rigidbody}. 
Interestingly, the two terms in~\eqref{finalErigid} become comparable for $J\sim Q^{3/2}$ which is another indication that the rigid body breaks down in that domain, as the energy of the background superfluid begins to be challenged by the potential of the vortices. (Another consistency check is to note that from \eqref{vortden} it follows that the number of vortices minus anti-vortices scales like $\sim J/Q$. On the other hand, since the area of each vortex is $\sim \mu^{-2}$, the number of vortices on the sphere is limited by $R^2\mu^2\sim Q$, which is indeed parametrically larger than $J/Q$.)

On general grounds, we expect that when the superfluid velocity exceeds the speed of sound $c_s=1/\sqrt{2}$ the rigid body configuration becomes unstable \cite{Bekenstein:1998nt}, and the superfluid settles in a new ground state \cite{fetter2005rapid}. Remarkably, such a transition was experimentally observed for Bose-Einstein condensates in anharmonic traps \cite{guo2020supersonic}.
 
\section{The Giant Vortex} 

\subsection{The Classical Profile}\label{subsec_GV1}

To guess the ground state past the point $J\sim Q^{3/2}$ is not obvious. Motivated from simulations of particles in a rotating trap and also from explicit solutions in classical field theories, we propose that an interesting next ground state is the giant vortex. 

The hallmark property of the giant vortex is that the superfluid only exists in a domain around the equator. The total vorticity is some $\ell\in \mathbb{Z}$ and, in the simplest case, which is the case we analyze here, inside the domain of the superfluid there are no additional vortices. The domain can now rotate supersonically compared to the velocity $1/\sqrt2$ of phonons in the original superfluid. This leads to no instabilities because the velocity of phonons in the giant vortex is no longer $1/\sqrt2$ (rather, the phonons can now move at the speed of light, as we will see in sec. \ref{subsec_fluct}).

Going back all the way to~\eqref{Hthe}, the equation of motion is 
\begin{equation}
\partial_\nu(\sqrt g g^{\mu\nu}|\partial \chi|\partial_\mu\chi) =0\,.
\end{equation}
An interesting solution with the required vorticity is 
 \begin{equation}\label{giantvor}
 \chi=\mu t+\ell\phi~.
 \end{equation}
In the parlance of section \ref{sec_vortices} the profile \eqref{giantvor} corresponds to a vortex-antivortex pair at the opposite poles.  The difference however is that we now consider values of the vorticities $\ell\sim R\mu $, i.e. very large vorticity. We will show below that this corresponds to the regime $J\gg Q^{3/2}$.

Since $\ell \in\mathbb{Z}$, the configuration \eqref{giantvor} only describes states with $J=\ell Q$; in sec. \ref{subsec_fluct} we will see that states with some nearby values of the spin, $| J-\ell Q|\ll Q $, are obtained by considering fluctuations of the field. As in the previous section, it would be interesting to find the classical profile which generalizes \eqref{giantvor} to arbitrary values of $J$.

\begin{figure}[t]
\begin{center}
\includegraphics[scale=0.18,trim= 0 3cm 0 6cm]{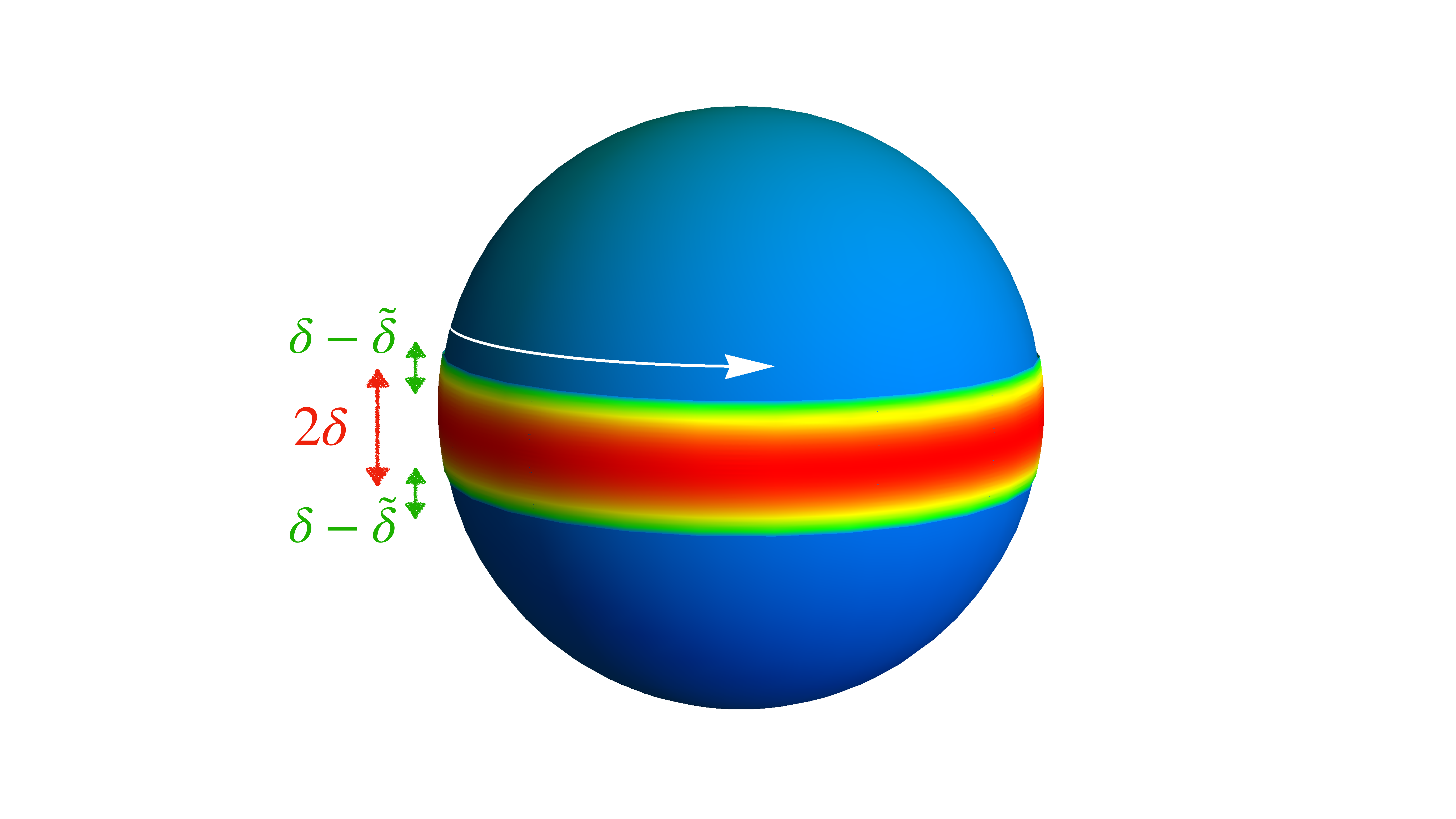}\caption{In the blue region there are no low-lying excitations, the red region is the spinning superfluid and the green region is the boundary layer, which we discuss in subsection~\ref{subsec_higher_der}. As long as the red region is large enough, we can trust the giant vortex effective theory.}\label{GV}
\end{center}
\end{figure}

The solution \eqref{giantvor} does not make sense everywhere on the sphere.  From the point of view of the superfluid effective theory we can continue trusting the configuration~\eqref{giantvor} for as long as $(\partial\chi)^2>0$ (since $(\partial\chi)^2$ controls the gap of the radial modes). This means the domain is given by $\theta\in [\pi/2-\delta,\pi/2+\delta]$ with $\cos \delta = |\ell|/( R\mu)$.
This also determines the upper limit on $\ell$ where the domain becomes very narrow ($\delta\ll1$).  The fundamental limitation comes from requiring that the domain over which the effective field theory is defined is larger than the inverse cutoff. See fig.~\ref{GV} (where also a boundary layer domain which we will soon discuss is depicted).
The size of the domain in the $\theta$ direction is $\sim R\delta$ while the energy cutoff, as inferred from the equator,  is $|\partial\chi|\sim\sqrt{\mu^2-\ell^2/R^2} \sim \mu\delta$.  
We hence obtain the inequality $\mu R\delta^2\gg  1$. We will now see what this implies for the regime of $Q,J$ that is achievable with the giant vortex effective theory.

In appendix~\ref{GiantVoQN} we compute the charge, angular momentum and energy for the solution \eqref{giantvor}. We find that the parameters $\mu$ and $\delta$ admit a simple expansion in $Q^3/ J^2 \ll 1$:
\begin{align}\label{eq_GV_mu}
R\mu &=\frac{J}{Q}\left[1+\frac{Q^3}{6\pi^2\alpha J^2}+
\ldots\right]\,,\\ \label{eq_GV_delta}
\sin(\delta) &=\frac{Q^{3/2}}{\sqrt{3\alpha}\pi J}\left[1-\frac{Q^3}{16\pi^2\alpha J^2}+
\ldots\right]\,.
\end{align}
Notice the size of the strip $\delta\sim Q^{3/2}/J$.  For $\delta\ll1$ the giant vortex allows us to reach angular momentum with $J\gg Q^{3/2}$. However, as discussed above, there is a limit to how small we can take the width of the strip, $\sim \delta$, to be. Our previous considerations have shown that $\delta$ is limited by $\sim 1/\sqrt{\mu R}$. Using eq.~\eqref{eq_GV_mu} we find that the largest angular momentum compatible with the giant vortex effective theory is $J\ll Q^2$. Finally we quote the equation of state
\begin{equation}\label{eqstate}
\Delta=RE=J+{1\over 12\pi^2\alpha} {Q^3\over J}+\cdots~,\quad Q^{3/2}\ll J\ll Q^2~.
\end{equation}

The equation of state has two noteworthy features:
\begin{itemize}
\item To leading order, the relationship between the energy and the angular momentum is $E={1\over R} J$. That the coefficient in this formula is exactly 1 is very important as this means that in terms of the scaling dimension~\eqref{radquan} we have $\Delta=J$ to leading order, which is precisely the slope also when the spin is the largest parameter. In other words, this is exactly the Regge slope!  
\item The subleading correction of order $Q^3/J$ becomes of order $Q$ near the boundary of the regime of validity of the giant vortex $J\sim Q^2$. This agrees parametrically with the sub-leading correction in the large-spin expansion, as we will review. Later we will see a striking resemblance between the fluctuations of the giant vortex and the fluctuations in the large-spin expansion. 
\end{itemize}

We stress that, even though eq.s \eqref{eq_GV_mu}, \eqref{eq_GV_delta} and \eqref{eqstate} are expressed as a series expansion, so far we worked only with the leading order action \eqref{Hthe}. Higher derivative corrections introduce correction suppressed by powers of $1/(\mu R\delta^2)^2\sim J^2/Q^4$. Here and in the following we work in the formal limit where $J/Q^2\rightarrow 0$ while $Q^3/J^2$ is fixed and small, so that we can safely neglect higher derivative terms while retaining the second term in eq. \eqref{eqstate}.  We return to this point in sec. \ref{subsec_higher_der}, where we provide a more careful analysis of the neglected contributions. 

A related comment is that it is useful to think about $M^2=-(\partial\chi)^2$ as an effective mass for a radial mode.\footnote{This is manifest in the linear sigma models considered, e.g., in \cite{Alvarez-Gaume:2016vff,Badel:2019khk}.} Outside the strip the theory is completely massive (which is why we can ignore the physics outside of the strip) and inside the strip the radial mode has nonzero expectation value and thus we have a superfluid theory.  The superfluid effective theory is cutoff at points $\theta=\pi/2\pm\tilde{\delta}$ at some distance from the (small) layer where $M^2\approx 0$. At those points we impose boundary conditions.  For the moment, it suffices to say that to leading order we can identify the location of the thickened boundary $\tilde{\delta}$ with $\delta$, and that Neumann boundary conditions  $j_\theta=0$ are imposed at that point independently of the microscopic model according to the general analysis in \cite{Cuomo:2021cnb}.
We discuss the physics of the thickened boundary more in detail in sec. \ref{subsec_higher_der}, where we show that the boundary layer leads to corrections to eq. \eqref{eqstate} which are suppressed by fractional powers of  $1/(\mu R\delta^2)^2\sim J^2/Q^4$.

\subsection{Fluctuations Around the Giant Vortex}\label{subsec_fluct}

Let us analyze fluctuations around the giant vortex. We simply expand around~\eqref{giantvor} with fluctuations denoted by $\varphi$ as in the previous section. We retain only the quadratic terms in the fluctuations (disposing of the linear, total derivative, terms and also of higher-order terms). We find the action for the fluctuations (assuming $\ell>0$ without loss of generality),
 \begin{equation}\label{flucLag}
 S_{{\rm fluctuations}}={3\over 2}\alpha\mu R^2\int dt \sin\theta d\theta d\phi\left[{1\over X}\left( \partial_0\varphi-{\cos\delta\over R\sin\theta}\partial_\phi\varphi\right)^2+X(\partial \varphi)^2\right]~.
\end{equation}
We denoted $X=\sqrt{1-{\cos^2\delta\over \sin^2\theta}}$ for simplicity. It is important to keep in mind that $\theta\in[\pi/2-\delta,\pi/2+\delta]$. More precisely, as discussed above, we must cut off the effective theory slightly before we hit $\theta=\pi/2\pm \delta$ and replace the singular boundary with a thickened boundary with Neumann boundary conditions $j_\theta\bigr|_{{\rm boundary}}=X\partial_\theta  \varphi=0$~\cite{Cuomo:2021cnb}.

The solutions to the equations of motion derived from~\eqref{flucLag} are not known to us in general, but a lot of the physics can be understood from the narrow strip $\delta\to0$ limit, in which the equations simplify.  Below we discuss in detail the result to leading order in $\delta$, and then quote some subleading orders.

In the limit $\delta\to 0$ we can replace the measure on the sphere $\sin\theta d\theta d\phi$ by just $d\theta d\phi$, which is consistent since the strip is a narrow band with opening angle $\sim Q^{3/2}/ J$ around the equator. Additionally,  we can approximate $\cos\delta\simeq 1$.
Finally, we see that everywhere within the strip of the superfluid we have $X\sim \delta$ or smaller, therefore, in the term $X(\partial\varphi)^2$ the time derivative and $\phi$ derivatives are negligible while the $\theta$ derivative should be retained since $\partial_\theta\sim 1/\delta$. 
We therefore get the theory for the fluctuations 
 \begin{equation}\label{flucLagi}
 S_{{\rm fluctuations}}\simeq{3\over 2}\alpha\mu R^2\int dt d\theta d\phi\left[\frac{1}{X}\left( \partial_0\varphi-{1\over R}\partial_\phi\varphi\right)^2
-\frac{X}{R^2}(\partial_\theta \varphi)^2  \right]~.
 \end{equation}
 The modes of the fluctuations can be found by doing a change of variables $d\theta/dy =X$, in terms of which we have the action 
\begin{equation}\label{flucLagii}
S_{{\rm fluctuations}}={3\over 2}\alpha\mu R^2\int dt dy d\phi\left[
\left( \partial_0\varphi-{1\over R}\partial_\phi\varphi\right)^2  
  -{1\over R^2}(\partial_y\varphi)^2\right]~.
  \end{equation}
 and the coordinate $y$ ranges in the interval $y\in[-\pi/2,\pi/2]$.\footnote{The solution of $d\theta/dy =X$ is 
 \begin{equation}
 y=-\arcsin\left({\cos\theta\over \sin\delta}\right)
\end{equation} 
and therefore as $\theta$ varies between $\theta=\pi/2-\delta$ to $\theta=\pi/2+\delta$, the variable $y$ varies between $y=-\pi/2$ to $y=\pi/2$.}
The boundary condition in these coordinates is naturally $\partial_y\varphi\bigr|_{y=\pm\pi/2}=0$.

Unlike the fluctuations around the non-rotating superfluid phase~\eqref{fluc}, the theory for the fluctuations~\eqref{flucLagii} is somewhat unfamiliar. 
The Neumann boundary conditions allow modes of the type 
$\cos( n y)$ for even $n$ or $\sin(n y)$ for odd $n$.  Therefore expanding with these eigenfunctions we have $\partial_y^2 \varphi = - n^2\varphi$ where $n$ is a positive integer (eigenfunctions with negative $n$ are redundant).  Therefore the dispersion relation obtained from the action~\eqref{flucLagii} is 
\begin{equation}\label{disrel}
\left(\omega-\frac{m}{R}\right)^2-\frac{n^2}{R^2}=0~,
\end{equation} 
with integer $m,n$ ($n\geq 0$). The corresponding one-particle states have angular momentum $J_z=\ell Q+m$, where $\ell Q$ is the contribution from the background.

This dispersion relation leads to a Fock space with some remarkable properties
\begin{itemize}
\item For $n=0$ we have a second order pole in the Green's function at every $\omega=m/R$.
The second order pole gets resolved at higher order in $\delta$ into two simple poles, of which one has to be interpreted as creation $a^\dagger_m$ and the other as annihilation operator $a_m$. The existence of creation operators with $\omega=m/R$ for any integer $m$ leads to an infinite degeneracy of the vacuum due to states of the form $\prod_i a_{m_i}^\dagger|vac\rangle$ with $\sum_i m_i=0$. This degeneracy also gets resolved at higher order in $\delta$, but for now, this is a peculiar property we have to keep in mind.\footnote{We discuss the quantization of the $n=0$ modes including the first subleading correction in appendix \ref{app_fluct}.}
Importantly, excitations with quantum number $m$ carry no $U(1)$ charge but they carry $m$ units of angular momentum $J_z$. Therefore states with $\sum m_i\neq 0$ either increase or decrease the total angular momentum, and correspondingly, increase or decrease the energy.
\item For $n\neq 0$ we have to keep only the positive square root of~\eqref{disrel} and we find 
\begin{equation}
\omega=m/R+n/R~,\qquad
n>0\,.
\end{equation} 
The fact that the integers $m,n$ appear with coefficient $1/R$ is remarkable -- we will see that this is essentially identical to the Fock space one obtains for multi-twist operators in the large spin limit.
\item As we discuss below, there are small $O(\delta)$ corrections to the dispersion relation, lifting the infinite degeneracy. However, some modes are protected by symmetry. The mode $m=-1,n=1$ is the soft mode ($\omega=0$) corresponding to rotations $J_-=J_+^\dagger=J_x-iJ_y$. This acts on our highest weight state and decreases the $z$ component of the angular momentum while  keeping the scaling dimension intact.  The energy of this mode remains exactly 0 for all $\delta$.
The conformal descendants are interpolated by the  $\omega = 1/R$ modes with $(m=-1,n=2)$, $(m=1,n=0)$, and $(m=0,n=1)$, which all together form a triplet corresponding to the three possible descendants obtained acting with the momentum generator $P_\mu$. These again receive no corrections to their dispersion relation $\omega=1/R$. Finally the $m=n=0$ mode needs to be treated separately as in the discussion in sec. \ref{sec_largeQ}.
\end{itemize}

We close this section by giving the first nontrivial corrections in $\delta$. These are straightforwardly derived by restoring the subleading orders in the action \eqref{flucLagii} and solving the equations of motion perturbatively in $\delta$. The result takes a qualitatively different form for $n\geq 2$, $n=1$, and $n=0$.
\begin{itemize}
\item for $n\geq 2$ the dispersion relation is corrected at order $O(\delta^2)$ and reads
\begin{equation}
R\omega=m+n-\frac{\delta^2}{4}(2m+n)+\ldots\,,\qquad
\text{for }n\geq 2\,.
\end{equation}
\item For $n=1$ the fist nontrivial correction arises at order $O(\delta^4)$, giving
\begin{equation}
R\omega=m+1-\frac{\delta^4}{16}(m^2+m)+\ldots\,,\qquad
\text{for }n=1\,.
\end{equation}
\item Finally, the dispersion relation for the $n=0$ mode differs at subleading orders between $m>0$ and $m<0$, giving
\begin{equation}\label{eq_fluct_0subleading}
R\omega=\begin{cases}\displaystyle
m-\frac{\delta^6}{512}(m^3-m)+\ldots & \text{for }m>0 \\[0.6em]
\displaystyle
m-\delta^2 m +\ldots& \text{for }m<0\,,
\end{cases}\qquad \text{for }n=0\,,
\end{equation}
where we reported the first nontrivial correction both for $m>0$ and $m<0$.
\end{itemize}
The result \eqref{eq_fluct_0subleading} is particularly significant because it predicts the gap between the giant vortex and the next to lightest state with the same angular momentum. This is made of a $(m=-2,n=0)$ quantum and a $(m=2,n=0)$ one. The gap reads
\begin{equation}
\delta\Delta= 2\delta^2+\ldots =\frac{2 Q^3}{3\alpha \pi^2 J^2}+\ldots\,.
\end{equation}

Finally from the subleading terms in the expansion of \eqref{flucLag}, we infer that the neglected nonlinearities are suppressed as long as $|\omega-m|\ll \mu \delta^2$, $n\ll \mu \delta^2$ and $|\omega+m|\delta^2\ll \mu \delta^2$.  This asymmetric structure follows from the broken Lorentz invariance and the almost chiral nature of the theory. It implies that $n\ll Q^2/J$, while for the $n=0$ modes our analysis holds as long as $|m|\ll R\mu \sim J/Q$.

\subsection{Comparison with the Large-Spin Expansion}

\begin{figure}[t]
\begin{center}
\includegraphics[scale=0.4,trim= 0 0 0 0]{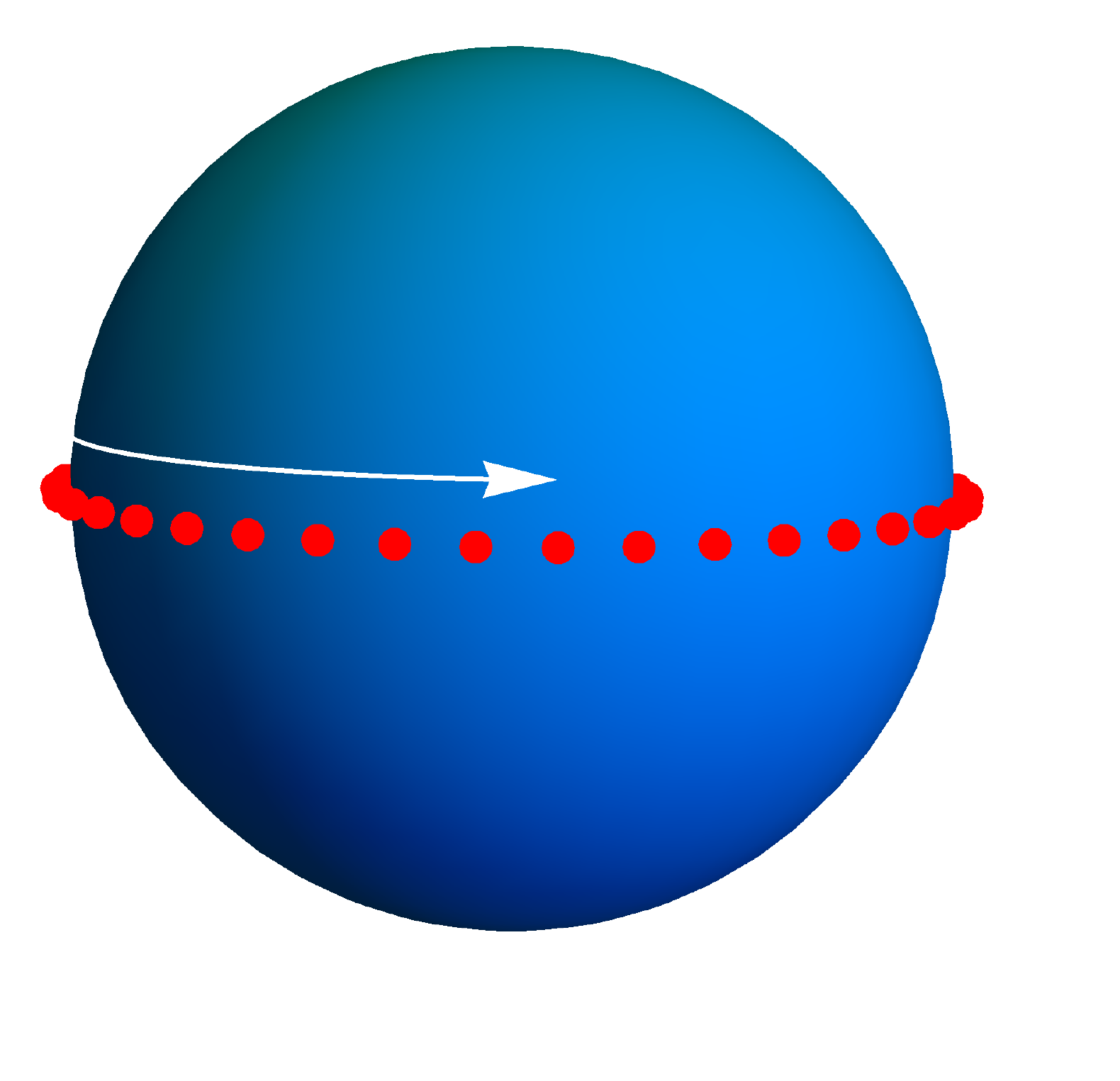}\caption{In the large spin limit, the minimal dimension operator(s) can be thought of as $Q$ partons rotating on the equator of the sphere. When $J/Q^2\sim 1$ these partons should collapse to the giant vortex.}\label{partons}
\end{center}
\end{figure}

Let us review the large-spin expansion and then compare with the giant vortex. In order to be concrete we will use notation appropriate for the $O(2)$ model, where there is an operator $\Phi$ carrying one unit of $U(1)$ charge. The lowest dimension operator of charge $Q$ is schematically 
$\Phi^Q$ and one now needs to add $J$ derivatives. Consider operators of the type 
\begin{equation}\label{schop}
\overset{\leftrightarrow}{ \partial^{a_1}}\Phi\overset{\leftrightarrow}{\partial^{a_2}}\Phi\cdots\overset{\leftrightarrow}{ \partial^{a_Q}}\Phi~,
\end{equation} 
 i.e. we insert many derivatives between any two $\Phi$ ``partons'' such that $\sum_{i=1}^Q a_i =J$.
If $J$ and each $a_i$ are large enough (we will soon compute how large they need to be) then the large-spin expansion applies and the leading and subleading order terms in the formula for the scaling dimension of the operator~\eqref{schop} are 
\begin{equation}\label{larges}
\Delta=J+Q\Delta_\Phi+\cdots~.
\end{equation}
One can loosely think about the corresponding state on $S^2$ in terms of $Q$ partons rotating on the equator of the sphere, see fig.~\ref{partons} 

Note that~\eqref{larges} implies a large degeneracy due to the insensitivity of $\Delta$ to rearrangements of the derivatives, to leading order. This is analogous to the large ground state degeneracy we encountered at leading order in $\delta$ in our analysis of the giant vortex. 
Next, derivatives with contracted indices can be added to the operator~\eqref{schop}  without changing the angular momentum.  
In fact, there are two ways to add contracted derivatives -- either in the form $\partial^\mu\Phi\cdots \partial_\mu\Phi$ which corresponds to $n=2$ since we have added two derivatives without changing the angular momentum or via $\partial^\nu\Phi\cdots \epsilon_{\mu \nu \rho} \partial^\rho \Phi$ which corresponds to $n=1$ since we have added  one derivative without changing the angular momentum. More generally, we can obtain all integer positive $n$ in this way, exactly matching the giant vortex fluctuations!

It is time to discuss the corrections to~\eqref{larges}, as this will allow us to determine when the large-spin expansion breaks down. The corrections to the scaling dimension~\eqref{larges} come from interactions between the $Q=1$ constitutes. The exchanged ``forces'' are due to $Q=0$ lowest-twist operators, where from the unit operator we obtain~\eqref{larges}. The next lowest twist operators in the $O(2)$ model are the energy momentum tensor and the $O(2)$ current, both of which have twist 1.  A qualitative estimate for the interaction between the $Q$ partons 
comes from thinking about the problem in the analog AdS$_4$ configuration of $Q$ partons spinning at radial distance $d\sim R_{AdS}\log(J/Q)$ from the center of AdS$_4$. The gravitational interaction between all the $\sim Q^2$ pairs thus scales as $\sim Q^3/J$.

We therefore see that the large-spin expansion breaks down at $J\sim Q^2$ and it is only valid for $J\gg Q^2$. This nicely matches with the boundary of the regime of validity of the giant vortex description and further reinforces the connection between the giant vortex theory and the large spin expansion. 

Conjecturally, the quantum theory of $Q$ partons with the $\sim Q^3/J$ interactions (mediated by the exchange of the energy momentum tensor and current) settles in a new ground 
state, which is essentially the giant vortex state.  This might be possible to demonstrate explicitly.  Qualitatively, since the interaction per parton scales like $Q^2/J$, once $J\sim Q^2$ the interaction provides sufficient energy for partons to climb over the $O(1)$ repulsive barrier of parton recombination. Notice that this picture is nicely consistent with the interaction between the partons being attractive.\footnote{
Let us check that the interaction between the partons is attractive. 
Denote the two-point functions of the $U(1)$ current and energy-momentum tensor by 
\begin{equation}
\langle J_\mu J_\nu \rangle={\tau\over 16\pi^2}{I_{\mu\nu}\over x^{4}}~,\qquad \langle T_{\mu\nu}T_{\rho\sigma}\rangle={C_{T}\over 16\pi^2}{I_{\mu\nu;\sigma\rho}\over x^6}
\end{equation}
where $I_{\mu\nu}=\delta_{\mu\nu}-2{x_\mu x_\nu\over x^2}$ and $I_{\mu\nu;\sigma\rho}=\frac12\left(I_{\mu\sigma}I_{\nu\rho}+I_{\mu\rho}I_{\nu\sigma}\right)-{1\over 3}\delta_{\mu\nu}\delta_{\sigma\rho} $ (see~\cite{Osborn:1993cr}). For a free complex scalar field in 2+1 dimensions of charge $1$ we have 
$\tau=2$ and $C_T=3$. In the interacting theory, we have approximately~\cite{Chester:2019ifh} that $\tau=1.809$ and $C_T=2.832$. Finally we recall that the parton-parton interaction is attractive if~\cite{Komargodski:2012ek} 
\begin{equation}\label{bind}
{\Delta_\Phi\over \sqrt C_T} \geq {1\over \sqrt 6} {1\over \sqrt \tau}~.
\end{equation}
Using the above values of $C_T,\tau$, and $\Delta_\Phi=0.519$ we find that the left hand side evaluates to $0.308$ while the right hand side to $0.304$.   Since the inequality~\eqref{bind} is indeed satisfied the force between the partons is attractive. This is consistent with our physical picture.
We expect however that a weakly repulsive force might also be compatible with a superfluid state; it is indeed known that bosons with repulsive interactions still form a superfluid state for generic values of the angular velocity \cite{2008JPCM...20l3202V}.
Note that, along the lines of our intuition from AdS,  if we could increase $C_T$ indefinitely and thereby decrease the Newton constant, we would never have binding gravitational interactions. 
Notice that inequalities similar to eq. \eqref{bind} were analyzed in \cite{Nakayama:2015hga} in relation with the weak gravity conjecture \cite{Arkani-Hamed:2006emk} in holography.
} 

Our discussion so far assumed that we break up the original operator $\Phi^Q$ into $Q$ constitutes separated by $\sim J/Q$ derivatives from each other. This is how one arrives at~\eqref{larges}. Another option is to, say, have a double parton without derivatives $\Phi^2$ and then $Q-1$ $\Phi$ partons separated from it and from each other. This has scaling dimension 
\begin{equation}\label{largespini}
RE=J+(Q-2)\Delta_{Q=1}+\Delta_{Q=2}+\cdots
\end{equation}
which is larger than~\eqref{larges} since $\Delta_{Q=2}>2\Delta_{Q=1}$, which is known to be true in the $O(2)$ model.\footnote{More generally, $\Delta_{Q}$ is conjecturally a convex function (it certainly is for very small and very large $Q$)~\cite{Aharony:2021mpc} and hence the energy is always minimized by breaking up the original operator $\Phi^Q$ into a maximal number of partons.}
In the effective theory around the giant vortex such a rearrangement of derivatives corresponds to turning on oscillators with large $|m|$ and we do not know whether in the giant vortex theory such operators are indeed $O(1)$ heavier than the ground state or, perhaps, they are much heavier. 

\subsection{Higher Derivative Corrections and the Boundary Layer}\label{subsec_higher_der}

We finally discuss higher derivative corrections. Our discussion will focus on the equation of state~\eqref{eqstate}, but analogous comments apply to the predictions discussed in sec.~\ref{subsec_fluct} for fluctuations.

As explained in sec. \ref{subsec_GV1},  the EFT can be schematically understood depending on the value of $M^2\equiv-(\p\chi)^2$.  When $M^2>0$ the theory is completely gapped; because of the centrifugal contribution $\ell^2/\sin^2\theta$, $M^2$ is positive and large everywhere but in a small strip of size $\delta$ around the equator. In the strip where $M^2<0$ we have a superfluid phase with local cutoff $\Lambda = |\partial\chi|\sim \mu \delta$.  Using that derivatives scale as $\pd\sim 1/\delta$, higher derivative operators, such as those shown in eq. \eqref{eq_termsNLO},  are locally suppressed by inverse integer powers of $\pd^2/\Lambda^2\sim J^2/Q^4$ (the square is because of parity invariance) with respect to the leading order action.  

It remains to discuss the boundaries of the region $[\pi/2- \delta,\pi/2+\delta]$, where $M^2$ vanishes and the effective theory naively breaks down. Physically, this means that we must avoid going all the way to $\pi/2\pm\delta$, and instead consider some region around the boundary and replace it by some effective boundary.
To understand how this works, we notice that integrating out the radial mode is disallowed when $\partial M \sim M^2$.
This defines a smaller region than the original $\theta\in[\pi/2-\delta,\pi/2+\delta]$, where the superfluid effective theory with a boundary condition should be truly valid. The region where we can use effective field theory is therefore $\theta\in[\pi/2-\tilde\delta,\pi/2+\tilde\delta]$ with some $\tilde\delta$ slightly smaller than $\delta$. We can obtain the difference $\delta-\tilde\delta$ from the requirement that the mass gradient is not at the cutoff scale,  $\p (\p\chi)\sim(\p\chi)^2$; we find
\begin{equation}
\delta-\tilde\delta\sim {\delta\over (\mu \delta^2 R)^{2/3}}\sim \left(Q^2\over J\right)^{1/3}{1\over \sqrt Q}\,.
\end{equation}

 Therefore all we have to do is to cut off from the strip a domain of size parametrically larger than $R\left(Q^2\over J\right)^{1/3}{1\over \sqrt Q} $ -- this allows to define within this slightly smaller strip a consistent effective theory with boundary conditions. See fig.~\ref{GV}. The size of the original strip is $2\delta\sim R Q^{3/2}/J$. Therefore the ratio between the boundary layer and the superfluid strip scales as 
\begin{equation}\label{eq_dmdb}
 (\delta-\tilde\delta)/\delta\sim \left(J\over Q^2\right)^{2/3}\,,
\end{equation} 
which is indeed a small number for $J\ll Q^2$.  The leading order boundary conditions at $\theta=\pi/2\pm\tilde{\delta}$ are found from the variation of the leading order bulk action \eqref{Hthe}, and simply correspond to Neumann $j_{\theta}=0$ \cite{Cuomo:2021cnb}.

To understand the consequences of this procedure of ``thickening the boundary'' for subleading terms in the derivative expansion, it is simplest to discuss the Lagrangian $L(\mu,\ell)$ evaluated on the profile \eqref{giantvor}, where $S=\int dt L$ is the total action given by the sum of \eqref{Hthe} and \eqref{eq_termsNLO}.\footnote{From the classical value of the Lagrangian $L(\mu,\ell)$, we can extract the $U(1)$ charge and the energy of the state as
\begin{equation}
Q=\frac{\p L}{\p\mu}\,,\qquad
\frac{\Delta}{R}=\mu Q-L\,,
\end{equation}
where the derivative is taken at fixed $\ell$.} The leading order action scales as $L_0=\alpha\int d^2x(\p\chi)^3\sim R^2\mu^3\delta^4\sim Q^3/J$.  Introducing a cutoff $\tilde{\delta}$ only changes the result of the integral by a term of order $\delta L_0\sim \mu^{4/3}\delta^{2/3}R^{1/3}\sim J^{2/3}/Q^{1/3}$.  Furthermore, introducing a cutoff $\tilde{\delta}$ is strictly necessary to compute the contribution $L_2$ from the subleading terms \eqref{eq_termsNLO}.  This is because, while they are locally $1/(R\mu\delta^2)^2$ suppressed at the equator, their expectation value $\sim \mu \delta^2/(\delta^2-\theta^2)^{3/2}$ is naively singular (and non-integrable) for $\theta\rightarrow\pi/2\pm \delta$. Integrating in the strip $[\pi/2-\tilde\delta,\pi/2+\tilde\delta]$ and using \eqref{eq_dmdb}, we find the schematic scaling 
\begin{equation}\label{eq_L2}
L_2\sim \mu^{4/3}\delta^{2/3} R^{1/3}\sim \frac{J^{2/3}}{Q^{1/3}}\,.
\end{equation}
This is the same scaling of $\delta L_0$ above. Notice that the precise value of $L_2$ depends upon the exact location of the cutoff $\tilde{\delta}$ and it is thus not calculable within EFT.
It may be similarly argued that also further higher derivative terms contribute to the same order because of their singular behaviour for $\theta\rightarrow \pi/2\pm\delta$.

We conclude that the boundary layer where $M^2\approx 0$ introduces a correction suppressed by the ratio $ L_2/L_0\sim 1/(\delta^2\mu R)^{5/3}\sim (J/Q^2)^{5/3}$ to $L$ and therefore to the equation of state \eqref{eqstate}.   Notice this correction is more important than that coming from the bulk of the strip, which is suppressed by $1/(\delta^2\mu R)^{2}\sim (J/Q^2)^{2}$.

As a final comment, we notice that the situation discussed above is analogous to that occurring for rotating strings in effective string theory \cite{Hellerman:2013kba} and non-relativistic superfluids in a harmonic trap \cite{Son:2005rv,Kravec:2018qnu}. Also in those cases, the contribution from infinitely many integrated higher derivative terms is enhanced with respect to their naive bulk scaling due to a singular behaviour at the boundary. It is known that in such setups to systematically compute subleading orders it is necessary to introduce a proper boundary action \cite{Hellerman:2016hnf,Hellerman:2020eff},  whose Wilson coefficients \emph{renormalize} the contributions from bulk operators and thus encode all the ambiguities related to boundary effects. We leave the explicit construction of such boundary action for future work.  Here we limit ourselves to noticing that the natural cutoff in the boundary layer is given by the inverse geometric size $\tilde{\Lambda}\sim R^{-1}/(\delta-\bar{\delta})$. Therefore, by locality, we expect the contribution of the boundary action to $L$ to scale as $\sim R\tilde{\Lambda}^2\sim \mu^{4/3}\delta^{2/3} R^{1/3}$, precisely matching eq. \eqref{eq_L2}; subleading terms should be further suppressed at least by powers of $\sim \pd^2/\tilde{\Lambda}^2\sim 1/(\mu\delta^2)^{4/3}\sim (J/Q^2)^{4/3} $. Therefore these considerations do not affect our analysis to the order of interest.

\section{Conclusions and Open Questions}

We have discussed the problem of determining the scaling dimensions of the lightest operators with large quantum numbers $Q,J$. This problem is equivalent to 
finding the phases of the theory at zero temperature and large $Q,J$ on $S^{d-1}$. Or, in the language of the grand-canonical ensemble, this is equivalent to finding the ground state of the zero temperature theory on $S^{d-1}$ with large chemical potentials for the $U(1)$ symmetry and angular momentum. The proposals we made in this paper should hold in the $O(2)$ model and other theories in 2+1 dimensions.

 We have seen that as the rotation increases, the superfluid develops vortices which at $J\sim Q$ form a rigid body, and when the rotation becomes supersonic at $J\sim Q^{3/2}$ presumably a large hole is created, to which we refer as the ``giant vortex.'' The giant vortex solution has remarkable properties as it allows to continue and increase the rotation further all the way to $J\sim Q^2$. Such fast rotation is allowed by a peculiar, chiral dispersion relation for the fluctuations of the giant vortex, $\omega={1\over R}\left(|n|+ m\right)$, allowing for the phonons to move at the speed of light. The fluctuations of the giant vortex lead to a Fock space which is virtually identical to the Fock space of multi-twist operators in the Regge limit, containing both the leading and the daughter Regge trajectories, therefore, allowing us to complete the phase diagram for all $Q,J$.

Several questions are left for the future.
\begin{itemize} 
\item It would be nice to explore further the phase diagram in 3+1 dimensions. There spinning operators are associated with vortex-strings; their configurations for sufficiently small $J$ have been studied in \cite{Cuomo:2019ejv}, but the analogue of the giant vortex is not known. Likewise, it would be nice to extend the discussion here to supersymmetric theories~\cite{Hellerman:2017veg,Hellerman:2017sur,Bourget:2018obm,Hellerman:2018xpi,Beccaria:2018xxl,Grassi:2019txd,Sharon:2020mjs,Hellerman:2021yqz} and to parity violating~\cite{Cuomo:2021qws} theories.\footnote{We also mention that spinning operators in NRCFTs were studied in \cite{Kravec:2019djc}.}
\item Starting from the large spin expansion, we have seen that if the energy-momentum and current exchanges could be solved exactly, it could be possible to derive the emergence of the giant vortex. In the AdS$_4$ analog of this question, this is a problem concerning the gravitational collapse of $Q$ charged particles.
\item Similarly, starting from the rigid body phase, it should be possible to follow the relativistic corrections and perhaps detect the (superradiant?) instability that leads to a new ground state. It should be possible to detect the instability in effective field theory, but to follow it through  one might need to start from a microscopic theory (Thomas-Fermi theory or the Gross–Pitaevskii model should suffice as well).
\item We do not presently know the orders of the various transitions we have described here, nor is it possible to be sure at the moment that there are no additional phases. Various phases and transitions with discrete symmetries have appeared in the context of particles in a trap; See~\cite{Garaud:2020qeq} for a recent discussion and references.  \item Some questions about the fluctuations of the giant vortex remain: What happens for large quantum numbers $m$? are the states described by moving elsewhere all the derivatives separating two adjacent partons gapped? 
How to describe giant vortices or the rigid body with $J$ not divisible by $Q$?
\end{itemize}

\section*{Acknowledgements}

We thank L. Delacr\'etaz, S. Dubovsky,  A. Hebbar, S.  Hellerman,  G.  Hern\'andez, A. Manenti, M. Metlitski,  B. Van Rees, and S.  Zare for useful discussions.  We are particularly grateful to O. Aharony, A.Monin, E. Palti, A. Raviv-Moshe and A. Sharon for providing comments on a preliminary version of this manuscript.  GC thanks the Galileo Galilei Institute,  \'Ecole Polytechnique and the organizers of the GGI workshop ``Bootstrapping Nature: Non-perturbative Approaches to Critical Phenomena"  for hospitality and support during the completion of this work.  GC is supported by the Simons Foundation (Simons Collaboration on the Non-perturbative Bootstrap) grants 488647 and 397411.  ZK is supported in part by the Simons Foundation grant 488657 (Simons Collaboration on the Non-Perturbative Bootstrap) and the BSF grant no. 2018204.  

\appendix

\section{Rigid Body rotation} 
\label{rigidbody}

We allow for electric sources in the EFT \eqref{Hthe} and look for the minimal energy configuration with a given angular momentum and charge. We look for azimuthally symmetric solutions with $\partial_\theta\chi=0$ and $\partial_\phi \chi=\mu R \sin\theta p(\theta)$ and $\partial_t \chi=\mu$.
This problem can be phrased as a search for solutions of azimuthally rotating superfluid (which inevitably leads to vortices if there is any $\theta$ dependence) with given angular momentum and charge. The time dependence is fixed by requiring that a diagonal subgroup involving time translations and the superfluid $U(1)$ symmetry is preserved.

To find the ground-state, we minimize the functional 
 \begin{equation}\label{functional}
E-\tilde{\mu}Q-\Omega J_z= 
 \mu^3\int d^2x\sqrt g \left(3 X^{1/2}- X^{3/2}-3\tilde \lambda X^{1/2}
-3\lambda  X^{1/2} p\sin\theta
 \right)~,
 \end{equation}
where $X=1-p^2$ and we set $\tilde{\mu}=\tilde{\lambda}\mu$ and $\Omega=\lambda/R$.
We get the equation 
\begin{equation}\label{azisuper}
p^3-\tilde\lambda p -\lambda \sin\theta\left(2p^2-1 \right)=0~.
\end{equation} 
Note the particular normalization of the Lagrange multipliers in~\eqref{functional} which renders the equations very convenient.
We should also minimize over the constant values of $\mu$ (at fixed $\tilde{\mu}$). We find the equation 
\begin{equation}\label{muminim}
\int d^2x\sqrt g X^{1/2} \left(3 - X-3\lambda  p\sin\theta-2\tilde \lambda \right)=0\quad\iff\quad
\frac{E-\Omega J_z}{Q}=\frac23\mu\tilde{\lambda}\,.
\end{equation}

\subsection{Reproducing the non-Relativistic Limit}

The non relativistic limit is small $\lambda$ and small $p$. Then~\eqref{azisuper} becomes a linear equation solved by 
\begin{equation}
p={\lambda\over\tilde \lambda} \sin\theta~.
\end{equation}
This is the rigid body solution discussed in the main text since $v^\phi=\partial^\phi \chi=const$.
To the first nontrivial order in $\lambda$ we can compute the quantum numbers: 
\begin{equation}
E=8\pi \mu^3R^2~,\quad J=8\pi {\lambda\over \tilde \lambda}\mu^3R^3 ~,\quad Q=12\pi \mu^2R^2-4\pi{\lambda^2\over \tilde\lambda^2} \mu^2R^2~.
\end{equation}
Note that the quadratic order in $\lambda$ cancels from $T_{00}$ and hence from $E$.
From these formulae we can extract the equation of state 
\begin{equation}
RE={1\over 3^{3/2}\sqrt\pi}Q^{3/2}+{3^{3/2}\sqrt \pi\over 2R}{J^2\over Q^{3/2}}\,.
\end{equation}
This exactly agrees with the results in the main text. Finally, we also need to impose~\eqref{muminim} which in the strict non-relativistic limit gives $\tilde \lambda=1$. Then, $\lambda$ is found from the angular momentum and $\mu$ is related to the charge, thereby fixing all the dimensionless unknowns in terms of physical quantities. 
From these formulas, we see that the relativistic limit corresponds to $\lambda/\tilde\lambda\sim1$, therefore we have $J\sim Q^{3/2}$ in the relativistic limit. 

We compute one more order in the non-relativistic limit. This already shows a deviation from the rigid body. The next-to-leading order solution is 
\begin{equation}\label{pform} 
p={\lambda\over \tilde \lambda}\sin\theta-{\lambda^3\over \tilde\lambda^3}\sin^3\theta\left(2-{1\over \tilde\lambda}\right)
\end{equation}
We can regard~\eqref{muminim} as an equation for $\tilde\lambda$, solving it, we find 
\begin{equation}
\tilde\lambda=1 -\frac23 \lambda^2~.
\end{equation}
For the velocity profile of the superfluid we just substitute this expression for $\tilde\lambda$ in~\eqref{pform} 
\begin{equation}
p=\lambda\left(1+{2\over 3}\lambda^2\right)\sin\theta-\lambda^3\sin^3\theta~.
\end{equation}
From current components  $j_0,j_\phi$ we can read out the velocity $j^\phi/j_0$ and the superfluid density $j_0$. We see that the angular velocity near the equator drops a little compared to the angular velocity elsewhere. Similarly, since ${1\over \sin\theta}\partial_\theta \partial_\phi\chi$ is the density of vortices, we see that it tends to decrease faster now as we approach the equator. In other words, vortices drift to the poles as the rotation velocity increases slightly. More generally, as we continue the expansion in the non relativistic limit, we will find a power series for $p$ of the form 
$p=\lambda\sin(\theta)\sum_nc_n(\lambda^2\sin^2\theta)^n$. 

Already for small $\lambda$ the cubic equation~\eqref{azisuper} admits additional solutions where $p$ does not go to zero as $\lambda\to0$. To leading order in $\lambda$ the two additional solutions are 
\begin{equation}\label{otherroots}
p=\pm \sqrt{\tilde \lambda}
+\lambda\left(1-{1\over 2\tilde\lambda}\right) \sin(\theta)~.
\end{equation}
We now have to impose~\eqref{muminim}, which to leading order in $\lambda$ gives $\tilde\lambda=2$, which is disallowed since then $X<0$ and the superfluid moves superluminaly. We therefore rule out the other two solutions of the cubic equation for small $\lambda$. It is tempting to conjecture that these additional solutions become viable in the relativistic regime and they play a role in the destabilization of the superfluid when it becomes supersonic. This is left for future work.

\section{The Charges of the Giant Vortex}\label{GiantVoQN}

In this appendix we compute the value of the conserved charges for the solution \eqref{giantvor}.
From the expression of the current \eqref{current}, we see that the charge is given by the following integral
\begin{equation}\label{eq_GV_Q}
 Q=6\pi R^2\alpha \mu^2 \int_{\pi/2-\delta}^{\pi/2+\delta} \sin \theta d\theta \sqrt{1-{\cos^2\delta\over \sin^2\theta} }~.
\end{equation} 
From the energy momentum tensor \eqref{emtensor} we find that the angular momentum is\footnote{Since $J_z=\ell Q$, one cannot truly cover all possible values of $Q^{3/2}\ll J\ll Q^2$ with the giant vortex -- only those that are approximately divisible by $Q$. We expect that a more general solution, which reduces to \eqref{giantvor} for $J_z=\ell Q$, might allow to describe all values  of $J_z$, but we were unable to find it.}  
\begin{equation}
J_z=\ell Q~,
\end{equation}
while the total energy of the giant vortex is 
\begin{equation}\label{eq_GV_E}
E=2\pi\alpha\mu^3 R^2 \int_{\pi/2-\delta}^{\pi/2+\delta}\sin\theta d\theta\left[ 3\sqrt{1-{\cos^2\delta\over \sin^2\theta} }-\left(1-{\cos^2\delta\over \sin^2\theta}\right)^{3/2}\right]\,.
\end{equation}

To evaluate eq.s \eqref{eq_GV_Q} and \eqref{eq_GV_E} explicitly, we recall the definition of the complete Elliptic integrals:
\begin{equation}
E(k)=\int_0^1 dx\frac{\sqrt{1-k^2 x^2}}{\sqrt{1-x^2}}\,,\qquad
K(k)=\int_0^1 dx\frac{1}{\sqrt{1-x^2}\sqrt{1-k^2 x^2}}\,.
\end{equation}
Upon performing the change of variable $\cos\theta=\sin\delta x$, it is possible to show that
\begin{align}
\int_{\pi/2-\delta}^{\pi/2+\delta} \sin \theta d\theta \sqrt{1-{\cos^2\delta\over \sin^2\theta} } &=
2 \left[E\left(\sin ^2(\delta) \right)-\cos ^2(\delta) \; K\left(\sin ^2(\delta )\right)\right]\,, \\[0.8em]
\int_{\pi/2-\delta}^{\pi/2+\delta} \sin \theta d\theta \left(1-{\cos^2\delta\over \sin^2\theta} \right)^{3/2}&=2\left\{\left[2-\sin^2(\delta)\right] E\left(\sin ^2(\delta )\right)-2 \cos ^2(\delta ) K\left(\sin ^2(\delta )\right)\right\}\,.
\end{align}
It follows that
\begin{align}\label{eq_GV_Q_full}
Q&=J_z/\ell=12 \pi  \alpha  \mu ^2 R^2\left[E\left(\sin ^2(\delta )\right)-\cos ^2(\delta ) K\left(\sin ^2(\delta )\right)\right]\,,\\[0.5em]
E&=
4 \pi  \alpha \mu ^3 R^2 \left[\left(1+ \sin ^2(\delta )\right) E\left(\sin ^2(\delta )\right)- \cos ^2(\delta ) K\left(\sin ^2(\delta )\right)\right]\,.
\label{eq_GV_E_full}
\end{align}
The expressions for the charge simply when expanded for small $\delta$, for which they read
\begin{align}\label{allchaii}
Q&=3\pi^2\alpha \mu^2R^2\left( \delta^2-{5\over 24}\delta^4 +\cdots\right)~, \\ \label{allchaiii} 
J_z&=3\pi^2\alpha \mu^3R^3\left(\delta^2-{17\over 24}\delta^4+\cdots\right)~,\\ \label{allchai}
E&=3\pi^2 \alpha\mu^3 R^2\left( \delta^2-{11\over 24}\delta^4+\cdots\right)~.
\end{align}

It is now simple to invert perturbatively eq.s \eqref{allchaii} and \eqref{allchaiii} to find $\mu$ and $\delta$ as a function of $Q$ and $J_z$.\footnote{In the opposite regime $\delta\simeq \pi/2$ one instead recovers the result for the energy discussed in sec.~\ref{sec_vortices} for point-like vortices.} The results are expressed as a series in $Q^3/(\alpha \pi^2 J^2)\ll 1$:
\begin{align}
\mu &=\frac{J}{Q}\left[1+\frac{Q^3}{6\pi^2\alpha J^2}+
O\left(\frac{Q^6}{\pi^4\alpha^2 J^4}\right)\right]\,,\\
\sin(\delta) &=\frac{Q^{3/2}}{\sqrt{3\alpha}\pi J}\left[1-\frac{Q^3}{16\pi^2\alpha J^2}+
O\left(\frac{Q^6}{\pi^4\alpha^2 J^4}\right)\right]\,.
\end{align}
Notice in particular that to leading order $\delta=  {1\over\pi\sqrt{3\alpha}}{Q^{3/2}\over J}$, which means that indeed the strip with the superfluid is small (i.e. it occupies a small fraction of there sphere) for $J\gg Q^{3/2}$. We can also infer the equation of state $E=E(Q,J)$: 
\begin{equation}
RE=J\left[1+{1\over 12\pi^2\alpha}{Q^3\over J^2}+
O\left(\frac{Q^6}{\pi^4\alpha^2 J^4}\right)\right]~.
\end{equation}

\section{Quantization of the \texorpdfstring{$n=0$}{n=0} Fluctuations of the Giant Vortex}\label{app_fluct}

To quantize the fluctuations with $n=0$ in the dispersion relation \eqref{disrel}, it is simplest to consider the effective theory which is obtained focusing on momenta $|\omega-m|\ll 1/R$, i.e. excluding all the $n\neq 0$ modes.  This amounts at considering only the quasi-homogeneous modes on the strip and thus the EFT is written in terms of a purely two-dimensional action. Denoting $\varphi_0$ the low energy (rescaled) field, we have
\begin{equation}\label{eq_0EFT2}
S_{\rm 0-modes}=R\int dt d\phi\left[\frac{1}{2}(\pd_-\varphi_0)^2+\frac{\delta^2}{4}\pd_+\varphi_0\pd_-\varphi_0+O\left(\delta^4\right)\right]\,,
\end{equation}
where we defined for brevity $\pd_{\mp}\varphi_0=\pd_0\varphi_0\mp\frac{1}{R}\pd_\phi \varphi_0$. 
The leading order is obtained simply upon taking $\varphi$ independent of $y$ in eq. \eqref{flucLagii}, while to obtain the subleading orders we need to match the result for the eigenfrequencies with the result obtained in the full theory.\footnote{To further subleading orders, higher derivative terms $(\pd^2\varphi_0)^2$ appear.} We need only consider the quadratic action for our purposes.
 
From eq. \eqref{eq_0EFT2} we obtain the mode expansion\footnote{Here we neglect the $m=0$ mode whose only role is to interpolate between states with different charge, see \cite{Cuomo:2020fsb,Dondi:2022wli} for details on its quantization.}
\begin{equation}
\begin{split}
\varphi_0 &\simeq
\sum_m e^{-i m t/R-im\phi}c_m+\sum_m e^{-i (m-\delta^2 m) t/R-im\phi}d_m \\
&=
\sum_m e^{-i mt/R-im\phi}\left(x_m+ p_m t/R\right)+O(\delta^2)\,,
\end{split}
\end{equation}
where the second line is the proper mode expansion in the $\delta\rightarrow 0$ limit. The $x_m$'s and $p_m$'s are then related to the $c_m$'s and $d_m$'s in the first line as
\begin{equation}\label{eq_app_c_d}
c_m=\frac12\left(x_m+i \frac{p_m}{m\delta^2/2}\right)\,,\qquad
d_m=\frac12\left(x_m-i \frac{p_m}{m\delta^2/2}\right)\,,
\end{equation}
The canonical momentum which follows from eq. \eqref{eq_0EFT2} is $p_0=\pd_+\varphi_0+O(\delta^2)$. Therefore the canonical commutation rules are
\begin{equation}
[\varphi_0(t,\phi),\pd_+\varphi_0(t,\phi')]=i\delta(\phi-\phi')/R\quad
\implies
\quad
[x_m,p_{-k}]=\frac{i}{2\pi}\delta_{n,k}\,,
\end{equation}
while $[x_m,x_k]=[p_m,p_k]=0$ $\forall \; m,k$.
Using eq.~\eqref{eq_app_c_d}, we obtain the following algebra for the operators $\{c_m,d_m\}$:
\begin{equation}
[c_m,c_{-m}]=\frac{1}{2\pi\delta^2 m}=-[d_m,d_{-m}]\,.
\end{equation}
We can therefore interpret $c_{m>0}$ and $d_{m<0}$ as destruction operators and define the vacuum as
\begin{equation}
c_m|0\rangle=d_{-m}|0\rangle=0\qquad
\forall\;m>0\,.
\end{equation}
The Fock space is the obtained acting on the vacuum with the operators $c_{m<0}$ and $d_{m>0}$. The dispersion relation of the single-particle states is given by
\begin{equation}
R\omega=\begin{cases}\displaystyle
m+O(\delta^4) & \text{for }m>0 \\[0.6em]
\displaystyle
m-\delta^2 m +O(\delta^4)& \text{for }m<0\,.
\end{cases}
\end{equation}
Upon including further subleading orders, we obtain eq. \eqref{eq_fluct_0subleading}.

\bibliography{Biblio}

\providecommand{\href}[2]{#2}\begingroup\raggedright\begin{thebibliography}{10}

\bibitem{Poland:2018epd}
D.~Poland, S.~Rychkov and A.~Vichi, \emph{{The Conformal Bootstrap: Theory,
  Numerical Techniques, and Applications}},
  \href{https://doi.org/10.1103/RevModPhys.91.015002}{\emph{Rev. Mod. Phys.}
  {\bfseries 91} (2019) 015002}
  [\href{https://arxiv.org/abs/1805.04405}{{\ttfamily 1805.04405}}].

\bibitem{Chester:2019wfx}
S.~M. Chester, \emph{{Weizmann Lectures on the Numerical Conformal Bootstrap}},
   \href{https://arxiv.org/abs/1907.05147}{{\ttfamily 1907.05147}}.

\bibitem{Hellerman:2015nra}
S.~Hellerman, D.~Orlando, S.~Reffert and M.~Watanabe, \emph{{On the CFT
  Operator Spectrum at Large Global Charge}},
  \href{https://doi.org/10.1007/JHEP12(2015)071}{\emph{JHEP} {\bfseries 12}
  (2015) 071} [\href{https://arxiv.org/abs/1505.01537}{{\ttfamily
  1505.01537}}].

\bibitem{Monin:2016jmo}
A.~Monin, D.~Pirtskhalava, R.~Rattazzi and F.~K. Seibold, \emph{{Semiclassics,
  Goldstone Bosons and CFT data}},
  \href{https://doi.org/10.1007/JHEP06(2017)011}{\emph{JHEP} {\bfseries 06}
  (2017) 011} [\href{https://arxiv.org/abs/1611.02912}{{\ttfamily
  1611.02912}}].

\bibitem{Gaume:2020bmp}
L.~A. Gaum\'e, D.~Orlando and S.~Reffert, \emph{{Selected topics in the large
  quantum number expansion}},
  \href{https://doi.org/10.1016/j.physrep.2021.08.001}{\emph{Phys. Rept.}
  {\bfseries 933} (2021) 1} [\href{https://arxiv.org/abs/2008.03308}{{\ttfamily
  2008.03308}}].

\bibitem{Jafferis:2017zna}
D.~Jafferis, B.~Mukhametzhanov and A.~Zhiboedov, \emph{{Conformal Bootstrap At
  Large Charge}}, \href{https://doi.org/10.1007/JHEP05(2018)043}{\emph{JHEP}
  {\bfseries 05} (2018) 043}
  [\href{https://arxiv.org/abs/1710.11161}{{\ttfamily 1710.11161}}].

\bibitem{Badel:2019oxl}
G.~Badel, G.~Cuomo, A.~Monin and R.~Rattazzi, \emph{{The Epsilon Expansion
  Meets Semiclassics}},
  \href{https://doi.org/10.1007/JHEP11(2019)110}{\emph{JHEP} {\bfseries 11}
  (2019) 110} [\href{https://arxiv.org/abs/1909.01269}{{\ttfamily
  1909.01269}}].

\bibitem{Alvarez-Gaume:2019biu}
L.~Alvarez-Gaume, D.~Orlando and S.~Reffert, \emph{{Large charge at large N}},
  \href{https://doi.org/10.1007/JHEP12(2019)142}{\emph{JHEP} {\bfseries 12}
  (2019) 142} [\href{https://arxiv.org/abs/1909.02571}{{\ttfamily
  1909.02571}}].

\bibitem{Banerjee:2017fcx}
D.~Banerjee, S.~Chandrasekharan and D.~Orlando, \emph{{Conformal dimensions via
  large charge expansion}},
  \href{https://doi.org/10.1103/PhysRevLett.120.061603}{\emph{Phys. Rev. Lett.}
  {\bfseries 120} (2018) 061603}
  [\href{https://arxiv.org/abs/1707.00711}{{\ttfamily 1707.00711}}].

\bibitem{Alday:2007mf}
L.~F. Alday and J.~M. Maldacena, \emph{{Comments on operators with large
  spin}}, \href{https://doi.org/10.1088/1126-6708/2007/11/019}{\emph{JHEP}
  {\bfseries 11} (2007) 019} [\href{https://arxiv.org/abs/0708.0672}{{\ttfamily
  0708.0672}}].

\bibitem{Komargodski:2012ek}
Z.~Komargodski and A.~Zhiboedov, \emph{{Convexity and Liberation at Large
  Spin}}, \href{https://doi.org/10.1007/JHEP11(2013)140}{\emph{JHEP} {\bfseries
  11} (2013) 140} [\href{https://arxiv.org/abs/1212.4103}{{\ttfamily
  1212.4103}}].

\bibitem{Fitzpatrick:2012yx}
A.~L. Fitzpatrick, J.~Kaplan, D.~Poland and D.~Simmons-Duffin, \emph{{The
  Analytic Bootstrap and AdS Superhorizon Locality}},
  \href{https://doi.org/10.1007/JHEP12(2013)004}{\emph{JHEP} {\bfseries 12}
  (2013) 004} [\href{https://arxiv.org/abs/1212.3616}{{\ttfamily 1212.3616}}].

\bibitem{Cuomo:2017vzg}
G.~Cuomo, A.~de~la Fuente, A.~Monin, D.~Pirtskhalava and R.~Rattazzi,
  \emph{{Rotating superfluids and spinning charged operators in conformal field
  theory}}, \href{https://doi.org/10.1103/PhysRevD.97.045012}{\emph{Phys. Rev.
  D} {\bfseries 97} (2018) 045012}
  [\href{https://arxiv.org/abs/1711.02108}{{\ttfamily 1711.02108}}].

\bibitem{Bekenstein:1998nt}
J.~D. Bekenstein and M.~Schiffer, \emph{{The Many faces of superradiance}},
  \href{https://doi.org/10.1103/PhysRevD.58.064014}{\emph{Phys. Rev. D}
  {\bfseries 58} (1998) 064014}
  [\href{https://arxiv.org/abs/gr-qc/9803033}{{\ttfamily gr-qc/9803033}}].

\bibitem{kasamatsu2002giant}
K.~{Kasamatsu}, M.~{Tsubota} and M.~{Ueda}, \emph{{Giant hole and circular
  superflow in a fast rotating Bose-Einstein condensate}},
  \href{https://doi.org/10.1103/PhysRevA.66.053606}{\emph{Physical Review A}
  {\bfseries 66} (2002) 053606}
  [\href{https://arxiv.org/abs/cond-mat/0202223}{{\ttfamily
  cond-mat/0202223}}].

\bibitem{Fischer:2003zz}
U.~R. Fischer and G.~Baym, \emph{{Vortex states of rapidly rotating dilute
  Bose-Einstein condensates}},
  \href{https://doi.org/10.1103/PhysRevLett.90.140402}{\emph{Phys. Rev. Lett.}
  {\bfseries 90} (2003) 140402}
  [\href{https://arxiv.org/abs/cond-mat/0111443}{{\ttfamily
  cond-mat/0111443}}].

\bibitem{fetter2005rapid}
A.~L. {Fetter}, B.~{Jackson} and S.~{Stringari}, \emph{{Rapid rotation of a
  Bose-Einstein condensate in a harmonic plus quartic trap}},
  \href{https://doi.org/10.1103/PhysRevA.71.013605}{\emph{Physical Review A}
  {\bfseries 71} (2005) 013605}
  [\href{https://arxiv.org/abs/cond-mat/0407119}{{\ttfamily
  cond-mat/0407119}}].

\bibitem{fu2006transition}
H.~{Fu} and E.~{Zaremba}, \emph{{Transition to the giant vortex state in a
  harmonic-plus-quartic trap}},
  \href{https://doi.org/10.1103/PhysRevA.73.013614}{\emph{Physical Review A}
  {\bfseries 73} (2006) 013614}
  [\href{https://arxiv.org/abs/cond-mat/0508515}{{\ttfamily
  cond-mat/0508515}}].

\bibitem{Fetter:2009zz}
A.~L. Fetter, \emph{{Rotating trapped Bose-Einstein condensates}},
  \href{https://doi.org/10.1103/RevModPhys.81.647}{\emph{Rev. Mod. Phys.}
  {\bfseries 81} (2009) 647}.

\bibitem{guo2020supersonic}
Y.~{Guo}, R.~{Dubessy}, M.~d.~G. {de Herve}, A.~{Kumar}, T.~{Badr}, A.~{Perrin}
  et~al., \emph{{Supersonic Rotation of a Superfluid: A Long-Lived Dynamical
  Ring}}, \href{https://doi.org/10.1103/PhysRevLett.124.025301}{\emph{Physical
  review letters} {\bfseries 124} (2020) 025301}
  [\href{https://arxiv.org/abs/1907.01795}{{\ttfamily 1907.01795}}].

\bibitem{Su:2022ysc}
J.-H. Su, C.-Y. Xia, W.-C. Yang and H.-B. Zeng, \emph{{Giant vortex in a fast
  rotating holographic superfluid}},
  \href{https://arxiv.org/abs/2208.14172}{{\ttfamily 2208.14172}}.

\bibitem{Penin:2020cxj}
A.~A. Penin and Q.~Weller, \emph{{What Becomes of Giant Vortices in the Abelian
  Higgs Model}},
  \href{https://doi.org/10.1103/PhysRevLett.125.251601}{\emph{Phys. Rev. Lett.}
  {\bfseries 125} (2020) 251601}
  [\href{https://arxiv.org/abs/2009.06640}{{\ttfamily 2009.06640}}].

\bibitem{Penin:2021xgr}
A.~A. Penin and Q.~Weller, \emph{{A theory of giant vortices}},
  \href{https://doi.org/10.1007/JHEP08(2021)056}{\emph{JHEP} {\bfseries 08}
  (2021) 056} [\href{https://arxiv.org/abs/2105.12137}{{\ttfamily
  2105.12137}}].

\bibitem{Fitzpatrick:2014vua}
A.~L. Fitzpatrick, J.~Kaplan and M.~T. Walters, \emph{{Universality of
  Long-Distance AdS Physics from the CFT Bootstrap}},
  \href{https://doi.org/10.1007/JHEP08(2014)145}{\emph{JHEP} {\bfseries 08}
  (2014) 145} [\href{https://arxiv.org/abs/1403.6829}{{\ttfamily 1403.6829}}].

\bibitem{Arkani-Hamed:2006emk}
N.~Arkani-Hamed, L.~Motl, A.~Nicolis and C.~Vafa, \emph{{The String landscape,
  black holes and gravity as the weakest force}},
  \href{https://doi.org/10.1088/1126-6708/2007/06/060}{\emph{JHEP} {\bfseries
  06} (2007) 060} [\href{https://arxiv.org/abs/hep-th/0601001}{{\ttfamily
  hep-th/0601001}}].

\bibitem{Palti:2019pca}
E.~Palti, \emph{{The Swampland: Introduction and Review}},
  \href{https://doi.org/10.1002/prop.201900037}{\emph{Fortsch. Phys.}
  {\bfseries 67} (2019) 1900037}
  [\href{https://arxiv.org/abs/1903.06239}{{\ttfamily 1903.06239}}].

\bibitem{Li:2015rfa}
D.~Li, D.~Meltzer and D.~Poland, \emph{{Non-Abelian Binding Energies from the
  Lightcone Bootstrap}},
  \href{https://doi.org/10.1007/JHEP02(2016)149}{\emph{JHEP} {\bfseries 02}
  (2016) 149} [\href{https://arxiv.org/abs/1510.07044}{{\ttfamily
  1510.07044}}].

\bibitem{Liu:2020tpf}
J.~Liu, D.~Meltzer, D.~Poland and D.~Simmons-Duffin, \emph{{The Lorentzian
  inversion formula and the spectrum of the 3d O(2) CFT}},
  \href{https://doi.org/10.1007/JHEP09(2020)115}{\emph{JHEP} {\bfseries 09}
  (2020) 115} [\href{https://arxiv.org/abs/2007.07914}{{\ttfamily
  2007.07914}}].

\bibitem{Aharony:2021mpc}
O.~Aharony and E.~Palti, \emph{{Convexity of charged operators in CFTs and the
  weak gravity conjecture}},
  \href{https://doi.org/10.1103/PhysRevD.104.126005}{\emph{Phys. Rev. D}
  {\bfseries 104} (2021) 126005}
  [\href{https://arxiv.org/abs/2108.04594}{{\ttfamily 2108.04594}}].

\bibitem{Antipin:2021rsh}
O.~Antipin, J.~Bersini, F.~Sannino, Z.-W. Wang and C.~Zhang, \emph{{More on the
  weak gravity conjecture via convexity of charged operators}},
  \href{https://doi.org/10.1007/JHEP12(2021)204}{\emph{JHEP} {\bfseries 12}
  (2021) 204} [\href{https://arxiv.org/abs/2109.04946}{{\ttfamily
  2109.04946}}].

\bibitem{Moser:2021bes}
R.~Moser, D.~Orlando and S.~Reffert, \emph{{Convexity, large charge and the
  large-N phase diagram of the \ensuremath{\varphi}$^{4}$ theory}},
  \href{https://doi.org/10.1007/JHEP02(2022)152}{\emph{JHEP} {\bfseries 02}
  (2022) 152} [\href{https://arxiv.org/abs/2110.07617}{{\ttfamily
  2110.07617}}].

\bibitem{Palti:2022unw}
E.~Palti and A.~Sharon, \emph{{Convexity of charged operators in CFTs with
  multiple Abelian symmetries}},
  \href{https://doi.org/10.1007/JHEP09(2022)078}{\emph{JHEP} {\bfseries 09}
  (2022) 078} [\href{https://arxiv.org/abs/2206.06703}{{\ttfamily
  2206.06703}}].

\bibitem{Nicolis:2015sra}
A.~Nicolis, R.~Penco, F.~Piazza and R.~Rattazzi, \emph{{Zoology of condensed
  matter: Framids, ordinary stuff, extra-ordinary stuff}},
  \href{https://doi.org/10.1007/JHEP06(2015)155}{\emph{JHEP} {\bfseries 06}
  (2015) 155} [\href{https://arxiv.org/abs/1501.03845}{{\ttfamily
  1501.03845}}].

\bibitem{Alberte:2020eil}
L.~Alberte and A.~Nicolis, \emph{{Spontaneously broken boosts and the Goldstone
  continuum}}, \href{https://doi.org/10.1007/JHEP07(2020)076}{\emph{JHEP}
  {\bfseries 07} (2020) 076}
  [\href{https://arxiv.org/abs/2001.06024}{{\ttfamily 2001.06024}}].

\bibitem{Komargodski:2021zzy}
Z.~Komargodski, M.~Mezei, S.~Pal and A.~Raviv-Moshe, \emph{{Spontaneously
  broken boosts in CFTs}},
  \href{https://doi.org/10.1007/JHEP09(2021)064}{\emph{JHEP} {\bfseries 09}
  (2021) 064} [\href{https://arxiv.org/abs/2102.12583}{{\ttfamily
  2102.12583}}].

\bibitem{Creminelli:2022onn}
P.~Creminelli, O.~Janssen and L.~Senatore, \emph{{Positivity bounds on
  effective field theories with spontaneously broken Lorentz invariance}},
  \href{https://doi.org/10.1007/JHEP09(2022)201}{\emph{JHEP} {\bfseries 09}
  (2022) 201} [\href{https://arxiv.org/abs/2207.14224}{{\ttfamily
  2207.14224}}].

\bibitem{Georgi:1992dw}
H.~Georgi, \emph{{Generalized dimensional analysis}},
  \href{https://doi.org/10.1016/0370-2693(93)91728-6}{\emph{Phys. Lett. B}
  {\bfseries 298} (1993) 187}
  [\href{https://arxiv.org/abs/hep-ph/9207278}{{\ttfamily hep-ph/9207278}}].

\bibitem{delaFuente:2018qwv}
A.~De~La~Fuente, \emph{{The large charge expansion at large $N$}},
  \href{https://doi.org/10.1007/JHEP08(2018)041}{\emph{JHEP} {\bfseries 08}
  (2018) 041} [\href{https://arxiv.org/abs/1805.00501}{{\ttfamily
  1805.00501}}].

\bibitem{Firat}
G.~Badel, E.~Firat, A.~Monin and R.~Rattazzi, ``Work in progress - private
  communication.''

\bibitem{Badel:2022fya}
G.~Badel, A.~Monin and R.~Rattazzi, \emph{{Identifying Large Charge
  Operators}},  \href{https://arxiv.org/abs/2207.08919}{{\ttfamily
  2207.08919}}.

\bibitem{onsager1949statistical}
L.~Onsager, \emph{Statistical hydrodynamics}, {\emph{Il Nuovo Cimento
  (1943-1954)} {\bfseries 6} (1949) 279}.

\bibitem{feynman1955chapter}
R.~P. Feynman, \emph{Chapter ii application of quantum mechanics to liquid
  helium},  in \emph{Progress in low temperature physics}, vol.~1, pp.~17--53.
\newblock Elsevier, 1955.

\bibitem{castin1999bose}
Y.~{Castin} and R.~{Dum}, \emph{{Bose-Einstein condensates with vortices in
  rotating traps}}, \href{https://doi.org/10.1007/s100530050584}{\emph{European
  Physical Journal D} {\bfseries 7} (1999) 399}
  [\href{https://arxiv.org/abs/cond-mat/9906144}{{\ttfamily
  cond-mat/9906144}}].

\bibitem{aftalion2001vortices}
A.~{Aftalion} and Q.~{Du}, \emph{{Vortices in a rotating Bose-Einstein
  condensate: Critical angular velocities and energy diagrams in the
  Thomas-Fermi regime}},
  \href{https://doi.org/10.1103/PhysRevA.64.063603}{\emph{Physical Review A}
  {\bfseries 64} (2001) 063603}
  [\href{https://arxiv.org/abs/cond-mat/0103299}{{\ttfamily
  cond-mat/0103299}}].

\bibitem{aftalion2007vortices}
A.~Aftalion, \emph{Vortices in Bose-Einstein Condensates}, vol.~67. Springer
  Science \& Business Media, 2007.

\bibitem{Horn:2015zna}
B.~Horn, A.~Nicolis and R.~Penco, \emph{{Effective string theory for vortex
  lines in fluids and superfluids}},
  \href{https://doi.org/10.1007/JHEP10(2015)153}{\emph{JHEP} {\bfseries 10}
  (2015) 153} [\href{https://arxiv.org/abs/1507.05635}{{\ttfamily
  1507.05635}}].

\bibitem{integrateLandau1}
N.~Sivan and S.~Levit, \emph{Semiclassical quantization of interacting
  electrons in a strong magnetic field},
  \href{https://doi.org/10.1103/PhysRevB.46.2319}{\emph{Phys. Rev. B}
  {\bfseries 46} (1992) 2319}.

\bibitem{integrateLandau2}
A.~Entelis and S.~Levit, \emph{Quantum adiabatic expansion for dynamics in
  strong magnetic fields},
  \href{https://doi.org/10.1103/PhysRevLett.69.3001}{\emph{Phys. Rev. Lett.}
  {\bfseries 69} (1992) 3001}.

\bibitem{integrateLandau3}
T.~Tochishita, M.~Mizui and H.~Kuratsuji, \emph{Semiclassical quantization for
  the motion of the guiding center using the coherent state path integral},
  \href{https://doi.org/10.1016/0375-9601(96)00073-4}{\emph{Physics Letters A}
  {\bfseries 212} (1996) 304 }.

\bibitem{jackiw1}
G.~V. Dunne, R.~Jackiw and C.~A. Trugenberger, \emph{``topological''
  (chern-simons) quantum mechanics},
  \href{https://doi.org/10.1103/PhysRevD.41.661}{\emph{Phys. Rev. D} {\bfseries
  41} (1990) 661}.

\bibitem{jackiw2}
G.~Dunne and R.~Jackiw, \emph{``peierls substitution'' and chern-simons quantum
  mechanics}, \href{https://doi.org/10.1016/0920-5632(93)90376-H}{\emph{Nuclear
  Physics B - Proceedings Supplements} {\bfseries 33} (1993) 114 }.

\bibitem{Shnir:2005vvi}
Y.~M. Shnir, \emph{{Magnetic Monopoles}}, Text and Monographs in Physics.
  Springer, Berlin/Heidelberg, 2005,
  \href{https://doi.org/10.1007/3-540-29082-6}{10.1007/3-540-29082-6}.

\bibitem{Garaud:2020qeq}
J.~Garaud and A.~J. Niemi, \emph{{Poincar\'e index formula and analogy with the
  Kosterlitz-Thouless transition in a non-rotated cold atom Bose-Einstein
  condensate}}, \href{https://doi.org/10.1007/JHEP09(2022)154}{\emph{JHEP}
  {\bfseries 22} (2020) 154}
  [\href{https://arxiv.org/abs/2108.03155}{{\ttfamily 2108.03155}}].

\bibitem{Alvarez-Gaume:2016vff}
L.~Alvarez-Gaume, O.~Loukas, D.~Orlando and S.~Reffert, \emph{{Compensating
  strong coupling with large charge}},
  \href{https://doi.org/10.1007/JHEP04(2017)059}{\emph{JHEP} {\bfseries 04}
  (2017) 059} [\href{https://arxiv.org/abs/1610.04495}{{\ttfamily
  1610.04495}}].

\bibitem{Badel:2019khk}
G.~Badel, G.~Cuomo, A.~Monin and R.~Rattazzi, \emph{{Feynman diagrams and the
  large charge expansion in $3-\varepsilon$ dimensions}},
  \href{https://doi.org/10.1016/j.physletb.2020.135202}{\emph{Phys. Lett. B}
  {\bfseries 802} (2020) 135202}
  [\href{https://arxiv.org/abs/1911.08505}{{\ttfamily 1911.08505}}].

\bibitem{Cuomo:2021cnb}
G.~Cuomo, M.~Mezei and A.~Raviv-Moshe, \emph{{Boundary conformal field theory
  at large charge}}, \href{https://doi.org/10.1007/JHEP10(2021)143}{\emph{JHEP}
  {\bfseries 10} (2021) 143}
  [\href{https://arxiv.org/abs/2108.06579}{{\ttfamily 2108.06579}}].

\bibitem{Osborn:1993cr}
H.~Osborn and A.~C. Petkou, \emph{{Implications of conformal invariance in
  field theories for general dimensions}},
  \href{https://doi.org/10.1006/aphy.1994.1045}{\emph{Annals Phys.} {\bfseries
  231} (1994) 311} [\href{https://arxiv.org/abs/hep-th/9307010}{{\ttfamily
  hep-th/9307010}}].

\bibitem{Chester:2019ifh}
S.~M. Chester, W.~Landry, J.~Liu, D.~Poland, D.~Simmons-Duffin, N.~Su et~al.,
  \emph{{Carving out OPE space and precise $O(2)$ model critical exponents}},
  \href{https://doi.org/10.1007/JHEP06(2020)142}{\emph{JHEP} {\bfseries 06}
  (2020) 142} [\href{https://arxiv.org/abs/1912.03324}{{\ttfamily
  1912.03324}}].

\bibitem{2008JPCM...20l3202V}
S.~{Viefers}, \emph{{Quantum Hall physics in rotating Bose Einstein
  condensates}},
  \href{https://doi.org/10.1088/0953-8984/20/12/123202}{\emph{Journal of
  Physics Condensed Matter} {\bfseries 20} (2008) 123202}
  [\href{https://arxiv.org/abs/0801.4856}{{\ttfamily 0801.4856}}].

\bibitem{Nakayama:2015hga}
Y.~Nakayama and Y.~Nomura, \emph{{Weak gravity conjecture in the AdS/CFT
  correspondence}},
  \href{https://doi.org/10.1103/PhysRevD.92.126006}{\emph{Phys. Rev. D}
  {\bfseries 92} (2015) 126006}
  [\href{https://arxiv.org/abs/1509.01647}{{\ttfamily 1509.01647}}].

\bibitem{Hellerman:2013kba}
S.~Hellerman and I.~Swanson, \emph{{String Theory of the Regge Intercept}},
  \href{https://doi.org/10.1103/PhysRevLett.114.111601}{\emph{Phys. Rev. Lett.}
  {\bfseries 114} (2015) 111601}
  [\href{https://arxiv.org/abs/1312.0999}{{\ttfamily 1312.0999}}].

\bibitem{Son:2005rv}
D.~T. Son and M.~Wingate, \emph{{General coordinate invariance and conformal
  invariance in nonrelativistic physics: Unitary Fermi gas}},
  \href{https://doi.org/10.1016/j.aop.2005.11.001}{\emph{Annals Phys.}
  {\bfseries 321} (2006) 197}
  [\href{https://arxiv.org/abs/cond-mat/0509786}{{\ttfamily
  cond-mat/0509786}}].

\bibitem{Kravec:2018qnu}
S.~M. Kravec and S.~Pal, \emph{{Nonrelativistic Conformal Field Theories in the
  Large Charge Sector}},
  \href{https://doi.org/10.1007/JHEP02(2019)008}{\emph{JHEP} {\bfseries 02}
  (2019) 008} [\href{https://arxiv.org/abs/1809.08188}{{\ttfamily
  1809.08188}}].

\bibitem{Hellerman:2016hnf}
S.~Hellerman and I.~Swanson, \emph{{Boundary Operators in Effective String
  Theory}}, \href{https://doi.org/10.1007/JHEP04(2017)085}{\emph{JHEP}
  {\bfseries 04} (2017) 085}
  [\href{https://arxiv.org/abs/1609.01736}{{\ttfamily 1609.01736}}].

\bibitem{Hellerman:2020eff}
S.~Hellerman and I.~Swanson, \emph{{Droplet-Edge Operators in Nonrelativistic
  Conformal Field Theories}},
  \href{https://arxiv.org/abs/2010.07967}{{\ttfamily 2010.07967}}.

\bibitem{Cuomo:2019ejv}
G.~Cuomo, \emph{{Superfluids, vortices and spinning charged operators in 4d
  CFT}}, \href{https://doi.org/10.1007/JHEP02(2020)119}{\emph{JHEP} {\bfseries
  02} (2020) 119} [\href{https://arxiv.org/abs/1906.07283}{{\ttfamily
  1906.07283}}].

\bibitem{Hellerman:2017veg}
S.~Hellerman, S.~Maeda and M.~Watanabe, \emph{{Operator Dimensions from
  Moduli}}, \href{https://doi.org/10.1007/JHEP10(2017)089}{\emph{JHEP}
  {\bfseries 10} (2017) 089}
  [\href{https://arxiv.org/abs/1706.05743}{{\ttfamily 1706.05743}}].

\bibitem{Hellerman:2017sur}
S.~Hellerman and S.~Maeda, \emph{{On the Large $R$-charge Expansion in
  ${\mathcal N} = 2$ Superconformal Field Theories}},
  \href{https://doi.org/10.1007/JHEP12(2017)135}{\emph{JHEP} {\bfseries 12}
  (2017) 135} [\href{https://arxiv.org/abs/1710.07336}{{\ttfamily
  1710.07336}}].

\bibitem{Bourget:2018obm}
A.~Bourget, D.~Rodriguez-Gomez and J.~G. Russo, \emph{{A limit for large
  $R$-charge correlators in $\mathcal{N}=2$ theories}},
  \href{https://doi.org/10.1007/JHEP05(2018)074}{\emph{JHEP} {\bfseries 05}
  (2018) 074} [\href{https://arxiv.org/abs/1803.00580}{{\ttfamily
  1803.00580}}].

\bibitem{Hellerman:2018xpi}
S.~Hellerman, S.~Maeda, D.~Orlando, S.~Reffert and M.~Watanabe,
  \emph{{Universal correlation functions in rank 1 SCFTs}},
  \href{https://doi.org/10.1007/JHEP12(2019)047}{\emph{JHEP} {\bfseries 12}
  (2019) 047} [\href{https://arxiv.org/abs/1804.01535}{{\ttfamily
  1804.01535}}].

\bibitem{Beccaria:2018xxl}
M.~Beccaria, \emph{{On the large R-charge $ \mathcal{N} $ = 2 chiral
  correlators and the Toda equation}},
  \href{https://doi.org/10.1007/JHEP02(2019)009}{\emph{JHEP} {\bfseries 02}
  (2019) 009} [\href{https://arxiv.org/abs/1809.06280}{{\ttfamily
  1809.06280}}].

\bibitem{Grassi:2019txd}
A.~Grassi, Z.~Komargodski and L.~Tizzano, \emph{{Extremal correlators and
  random matrix theory}},
  \href{https://doi.org/10.1007/JHEP04(2021)214}{\emph{JHEP} {\bfseries 04}
  (2021) 214} [\href{https://arxiv.org/abs/1908.10306}{{\ttfamily
  1908.10306}}].

\bibitem{Sharon:2020mjs}
A.~Sharon and M.~Watanabe, \emph{{Transition of Large $R$-Charge Operators on a
  Conformal Manifold}},
  \href{https://doi.org/10.1007/JHEP01(2021)068}{\emph{JHEP} {\bfseries 01}
  (2021) 068} [\href{https://arxiv.org/abs/2008.01106}{{\ttfamily
  2008.01106}}].

\bibitem{Hellerman:2021yqz}
S.~Hellerman and D.~Orlando, \emph{{Large R-charge EFT correlators in N=2
  SQCD}},  \href{https://arxiv.org/abs/2103.05642}{{\ttfamily 2103.05642}}.

\bibitem{Cuomo:2021qws}
G.~Cuomo, L.~V. Delacretaz and U.~Mehta, \emph{{Large Charge Sector of 3d
  Parity-Violating CFTs}},
  \href{https://doi.org/10.1007/JHEP05(2021)115}{\emph{JHEP} {\bfseries 05}
  (2021) 115} [\href{https://arxiv.org/abs/2102.05046}{{\ttfamily
  2102.05046}}].

\bibitem{Kravec:2019djc}
S.~M. Kravec and S.~Pal, \emph{{The Spinful Large Charge Sector of
  Non-Relativistic CFTs: From Phonons to Vortex Crystals}},
  \href{https://doi.org/10.1007/JHEP05(2019)194}{\emph{JHEP} {\bfseries 05}
  (2019) 194} [\href{https://arxiv.org/abs/1904.05462}{{\ttfamily
  1904.05462}}].

\bibitem{Cuomo:2020fsb}
G.~F. Cuomo, \emph{{Large charge, semiclassics and superfluids: from broken
  symmetries to conformal field theories}}, Ph.D. thesis, EPFL, 2020.
\newblock 10.5075/epfl-thesis-8397.

\bibitem{Dondi:2022wli}
N.~Dondi, I.~Kalogerakis, R.~Moser, D.~Orlando and S.~Reffert, \emph{{Spinning
  correlators in large-charge CFTs}},
  \href{https://doi.org/10.1016/j.nuclphysb.2022.115928}{\emph{Nucl. Phys. B}
  {\bfseries 983} (2022) 115928}
  [\href{https://arxiv.org/abs/2203.12624}{{\ttfamily 2203.12624}}].

\end{thebibliography}\endgroup
	\bibliographystyle{JHEP.bst}

\end{document}